# Early Release Science of the exoplanet WASP-39b with JWST NIRSpec PRISM

Z. Rustamkulov[1*], D. K. Sing[1,2], S. Mukherjee[3], E. M. May[4], J. Kirk[5,6,7], E. Schlawin[8], M. R. Line[9], C. Piaulet[10], A. L. Carter[3], N. E. Batalha[11], J. M. Goyal[12], M. López-Morales[5], J. D. Lothringer[13], R. J. MacDonald[14,15,16], S. E. Moran[17], K. B. Stevenson[4], H. R. Wakeford[18], N. Espinoza[19], J. L. Bean[20], N. M. Batalha[3], B. Benneke[21], Z. K. Berta-Thompson[22], I. J. M. Crossfield[23], P. Gao[24], L. Kreidberg[25], D. K. Powell[26,27], P. E. Cubillos[28], N. P. Gibson[29], J. Leconte[30], K. Molaverdikhani[31,32], N. K. Nikolov[19], V. Parmentier[33,34], P. Roy[35], J. Taylor[36], J. D. Turner[37,38], P. J. Wheatley[39,40], K. Aggarwal[41], E. Ahrer[39,40], M. K. Alam[24], L. Alderson[18], N. H. Allen[2,42], A. Banerjee[43], S. Barat[44], D. Barrado[45], J. K. Barstow[43], T. J. Bell[46], J. Blecic[47,48], J. Brande[49], S. Casewell[50], Q. Changeat[51,19,52], K. L. Chubb[53], N. Crouzet[54], T. Daylan[55,56], L. Decin[57], J. Désert[44], T. Mikal-Evans[25], A. D. Feinstein[20,42], L. Flagg[58], J. J. Fortney[3], J. Harrington[59], K. Heng[31], Y. Hong[37], R. Hu[60,61], N. Iro[62], T. Kataria[63], E. M.-R. Kempton[64], J. Krick[65], M. Lendl[66], J. Lillo-Box[67], A. Louca[54], J. Lustig-Yaeger[4], L. Mancini[68,69,25], M. Mansfield[8,27], N. J. Mayne[70], Y. Miguel[54,71], G. Morello[72,73,74], K. Ohno[3], E. Palle[72], D. J. M. Petit dit de la Roche[66], B. V. Rackham[75,76,77], M. Radica[10], L. Ramos-Rosado[2], S. Redfield[78], L. K. Rogers[79], E. L. Shkolnik[9], J. Southworth[80], J. Teske[24], P. Tremblin[81], G. S. Tucker[82], O. Venot[83], W. C. Waalkes[84], L. Welbanks[9,27], X. Zhang[85], S. Zieba[25,54]

*Corresponding author's email: zafar@jhu.edu
All author affiliations are listed at the end of the paper

**Transmission spectroscopy[1,2,3] of exoplanets has revealed signatures of water vapor, aerosols, and alkali metals in a few dozen exoplanet atmospheres[4,5]. However, these previous inferences with the Hubble and Spitzer Space Telescopes were hindered by the observations' relatively narrow wavelength range and spectral resolving power, which precluded the unambiguous identification of other chemical species — in particular the primary carbon-bearing molecules[6,7]. Here we report a broad-wavelength 0.5–5.5 $\mu$m atmospheric transmission spectrum of WASP39 b[8], a 1200 K, roughly Saturn-mass, Jupiter-radius exoplanet, measured with JWST NIRSpec's PRISM mode[9] as part of the JWST Transiting Exoplanet Community Early Release Science Team program[10,11,12]. We robustly detect multiple chemical species at high significance, including Na (19$\sigma$), $H_2O$ (33$\sigma$), $CO_2$ (28$\sigma$), and CO (7$\sigma$). The non-detection of $CH_4$, combined with a strong $CO_2$ feature, favours atmospheric models with a super-solar atmospheric metallicity. An unanticipated absorption feature at 4$\mu$m is best explained by $SO_2$ (2.7$\sigma$), which could be a tracer of atmospheric photochemistry. These observations demonstrate JWST's sensitivity to a rich diversity of exoplanet compositions and chemical processes.**

We observed one transit of WASP-39b on 10 July 2022 with JWST's Near InfraRed Spectrograph (NIRSpec)[9,13], using the PRISM mode, as part of the JWST Transiting Exoplanet Community Early Release Science Program (ERS Program 1366) (PIs: N. Batalha,

J. Bean, K. Stevenson)[10,11]. These observations cover the 0.5–5.5$\mu$m wavelength range at a native resolving power of R = $\lambda/\Delta\lambda$ ~ 20–300. WASP-39b was selected for this JWST ERS Program due to previous space- and ground-based observations revealing strong alkali metal absorption and multiple prominent $H_2O$ bands[4,6,14,15,16], suggesting strong signal-to-noise could be obtained with JWST. However, the limited wavelength range of existing transmission spectra (0.3–1.65$\mu$m, combined with two wide photometric Spitzer channels at 3.6 and 4.5$\mu$m) left several important questions unresolved. Previous estimates of WASP-39b's atmospheric metallicity—a measure of the relative abundance of all gases heavier than hydrogen or helium—vary by four orders of magnitude[6,16,17,18,19,20]. Accurate determinations of metallicity can elucidate formation pathways and provide greater insight into the planet's history[21]. The JWST NIRSpec PRISM observations we present here offer a more detailed view into WASP-39b's atmospheric composition than has previously been possible (see ref. 21 for an initial infrared analysis of this data).

We obtained time-series spectroscopy over 8.23 hours centered around the transit event to extract the wavelength-dependent absorption by the planet's atmosphere—i.e., the transmission spectrum, which probes the planet's day-night terminator region near millibar pressures. We used NIRSpec PRISM in Bright Object Time Series (BOTS) mode. WASP-39 is a bright, nearby, relatively inactive[23] G7 type star with an effective temperature of 5400 K[8]. WASP-39's J-band magnitude of 10.66 puts it near PRISM's saturation limit, which fortuitously allows us to test the effects of saturation on the quality of the resulting science compared to past measurements (see Methods).

In our baseline reduction using FIREFLy Fast InfraRed Exoplanet Fitting for Lightcurves[24], we perform calibrations on the raw data using the jwst Python pipeline[12], and then identify and correct for bad pixels and cosmic rays. We mitigate the 1/$f$ noise[9] at the group level rather than the integration level to ensure accurate slope fitting, which we find to be a crucial step for NIRSpec PRISM observations with few groups per integration.

We bin the resulting spectrophotometry in wavelength to create 207 variable-width spectral channels with roughly equal counts in each. Fig. 1 shows the FIREFLy white and spectrophotometric light curves at this step in the top panel. Several absorption features are visible by-eye as darker horizontal stripes within the transit region in the 2D light curve (Fig. 1), demonstrating the high quality of the raw spectrophotometry achieved by the PRISM observing mode.

To extract the atmosphere's transmission spectrum, we fit the planet's transit depth in each wavelength bin using a limb darkened transit light curve model using the Python-based Levenburg-Marquardt least squares algorithm lmfit[25]. The light curves show a typical photometric scatter of 0.2–1.2% per integration (1.36 seconds each), and the typical transit depth uncertainties vary between 50–200 parts-per-million (ppm), which is in line with near-photon-limited precision (see Methods). While we successfully measure fluxes in the saturated regions (0.8-2.3 $\mu$m), due to the lower number of groups used per integration here (1-3), the measured count rates may be adversely affected. We do not find excess red noise in the saturated channels themselves, however we notice large point-to-point scatter in the transit depths, which required wider wavelength binning to better match previous HST observations. Fig. 2 highlights representative transit light curves spanning the entire wavelength range. These data are binned into wider wavelength channels than those used for the final transmission spectrum for ease of presentation. Light curve systematics have not been removed from these data, demonstrating the unprecedented stability and precision of the PRISM observing mode.

We also compared the results from the FIREFLy reduction to three other independent reductions that use different treatments for the saturated region of the detector, limb darkening, and various detector systematics (see Methods). All four reductions obtain consistent results. Fig. 3 shows a comparison of the four reductions. The consistency provides confidence in the accuracy of derived atmospheric parameters, demonstrating that any residual systematics are minimal and do not strongly bias results for NIRSpec PRISM observations. The transmission spectrum also agrees well with previous measurements from ground-based telescopes[15,16] as well as HST and Spitzer[6] within error (see Fig. 3), indicating that we can reliably recover a spectrum at these levels of saturation. These PRISM observations offer high-quality data from 0.5–5.5 $\mu$m, with minimal contributions from systematics, and at precisions generally near the photon limit (see Methods). While recovery of the saturated region (0.9–1.5$\mu$m) is possible, caution is warranted when interpreting this portion of the spectrum (see Methods). Future PRISM observations of similarly bright targets should therefore carefully consider if saturating the spectrum is an appropriate choice for a given planet, or if building the wavelength coverage from multiple transits with different complementary modes is preferable.

The transmission spectrum of WASP-39b from the FIREFLy reduction is shown in Fig. 4. We select the FIREFLy reduction to be our baseline reduction, but comparable results are achieved with the three other reductions presented in this work (see Methods). We interpret the spectrum with grids of one-dimensional (1D) radiative–convective–thermochemical–equilibrium models (post-processed with some additional gases (see the Methods)), with a representative best-fitting model transmission spectrum shown in Fig. 4, along with opacity contributions from atoms, molecules, and grey clouds. We detect the presence of $H_2O$ via four pronounced independent bands (33$\sigma$, 1–2.2 $\mu$m), a prominent $CO_2$ feature at 4.3 $\mu$m (28$\sigma$), Na at 0.58 $\mu$m (19$\sigma$), a CO absorption band at 4.7 $\mu$m (7$\sigma$), and a grey cloud (21$\sigma$). We do not observe any significant $CH_4$ absorption (expected at 3.3 $\mu$m), despite predictions of its presence for atmospheres at approximately solar metallicity and place an upper limit of / 5×10$^{-6}$ on the $CH_4$ volume mixing ratio between 0.1–2 mbars. We also observe a relatively narrow absorption feature at 4.05 $\mu$m (~2.7$\sigma$), which we attribute to $SO_2$ — a potential tracer for photochemistry[26,27,28] — after an extensive search across many possible opacity sources (see Methods). Using a Bayesian approach described in the methods section, we calculate that the volume mixing ratio of $SO_2$ needed to explain this feature is 10$^{-5}$. The potential $SO_2$ feature is also observed at higher resolutions with JWST NIRSpec G395H[29], adding confidence that the feature first reported as an unknown absorber[22] is a genuine feature of the planet's atmosphere. With Na detected in the atmosphere, the alkali metal, K, is also expected at optical wavelengths[14] though not detected. However, the resolution covering the narrow K absorption doublet in the optical is low, which may be preventing detection. This might also be because of detector saturation in the wavelength range where K absorption is expected. We also do not detect the presence of $H_2S$ in the atmosphere. We note that although the best-fitting models shown in Figs. 3 and 4 have some $CH_4$, $H_2S$, and K signatures, these species are not favored by the data to the level of a detection. We determine the single best-fitting atmospheric metallicity, C/O ratio, and grey cloud opacity to be 10×solar, 0.7, and $\kappa_{cld}=10^{-2.07}$ cm$^2$/g, respectively. A detailed discussion on these best-fitting parameters is presented in the methods section.

JWST/NIRSpec PRISM's power to constrain multiple chemical species in hot giant planet atmospheres provides new windows into their compositions and chemical processes, as we show here with WASP-39b. Using our model grids, we find that WASP-39 b's best-fitting atmospheric metallicity is ~10× solar. In the limit of equilibrium chemistry, our non-detection of $CH_4$ at 3.4 $\mu$m paired with the prominence of the large $CO_2$ feature at 4.4 $\mu$m are indicative of a super-solar atmospheric metallicity, as illustrated in Fig. 13. This may point to WASP-39 b's puffy envelope bearing more compositional similarity to the similarly massed ice giants than the gas giants. Moreover, the likely detection of $SO_2$, and its unexpectedly high estimated abundance, suggests that photochemical processes are pushing this species out of equilibrium. Photochemistry models show that sulphur compounds such as $H_2S$ efficiently photodissociate and recombine to form $SO_2$ with ~1 ppm abundances and at 1-100 mbar pressures[27]—roughly the same pressure range probed by our transmission spectroscopy (see Fig.14). The abundance measurement of $SO_2$ can therefore serve as an important tracer of the thermochemical properties of highly irradiated stratospheres and the efficiency of photochemistry. Furthermore, our detection of a qualitatively significant wavelength dependence to the planet's central transit time (Fig. 7) suggests that these observations are sensitive to differences in the atmospheric composition at the planet's leading and trailing hemispheres. The measured ~20 second amplitude of this effect is in-line with model expectations[30]. This indicates that such observations will be informative in exploring the 3D nature of hot Jupiter atmospheres, which may give a more holistic understanding of their heat redistribution and nightside chemistry.

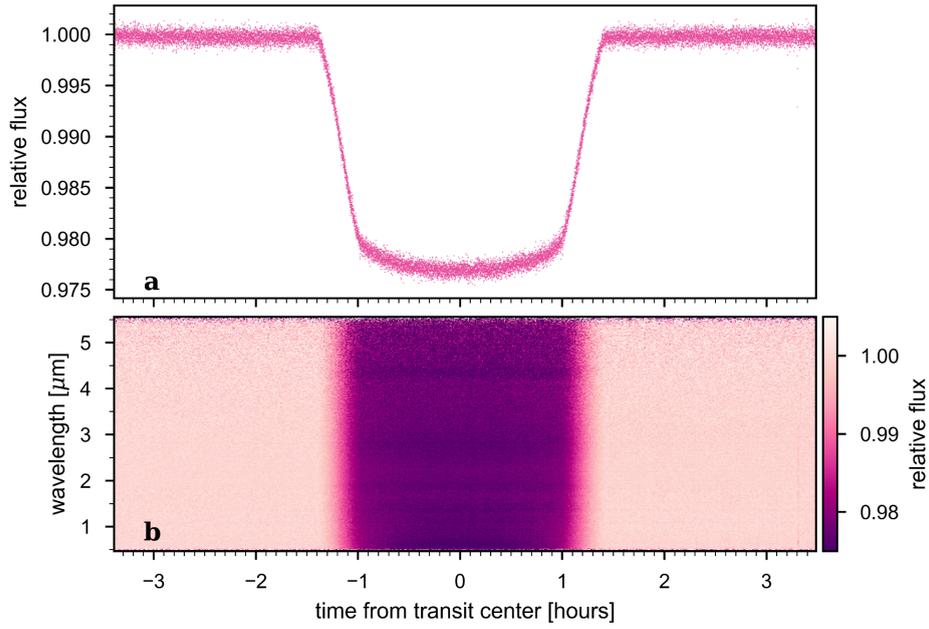

**Figure 1: The light curve of WASP-39b observed by JWST NIRSpec PRISM. a,** The normalized white light curve created by integrating over all wavelengths using the FIREFLy reduction. **b,** The binned time-series (with 30 integrations per time bin) of the relative flux for each wavelength. A constant 200 ppm/hour linear trend through time has been removed from the white light curve and each spectral channel for visual clarity.

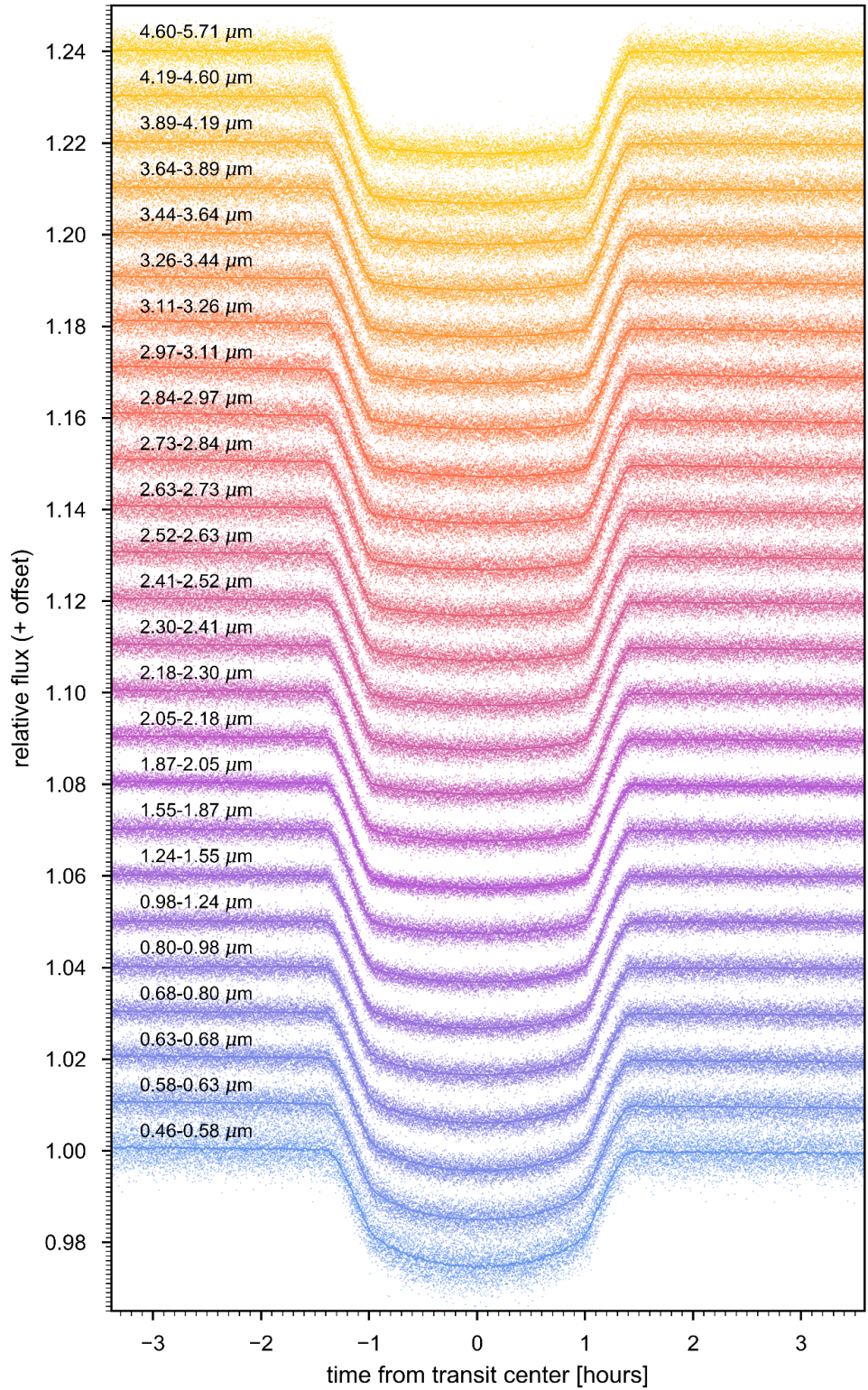

**Figure 2: Normalized spectrophotometric light curves for the JWST-PRISM transit of WASP-39b.** The light curves were created by summing over wide wavelength channels (wavelength ranges indicated on the plot). Overplotted on each light curve are their best-fit models, which include a transit model and detector systematics. Light curve systematics have not been removed from the data.

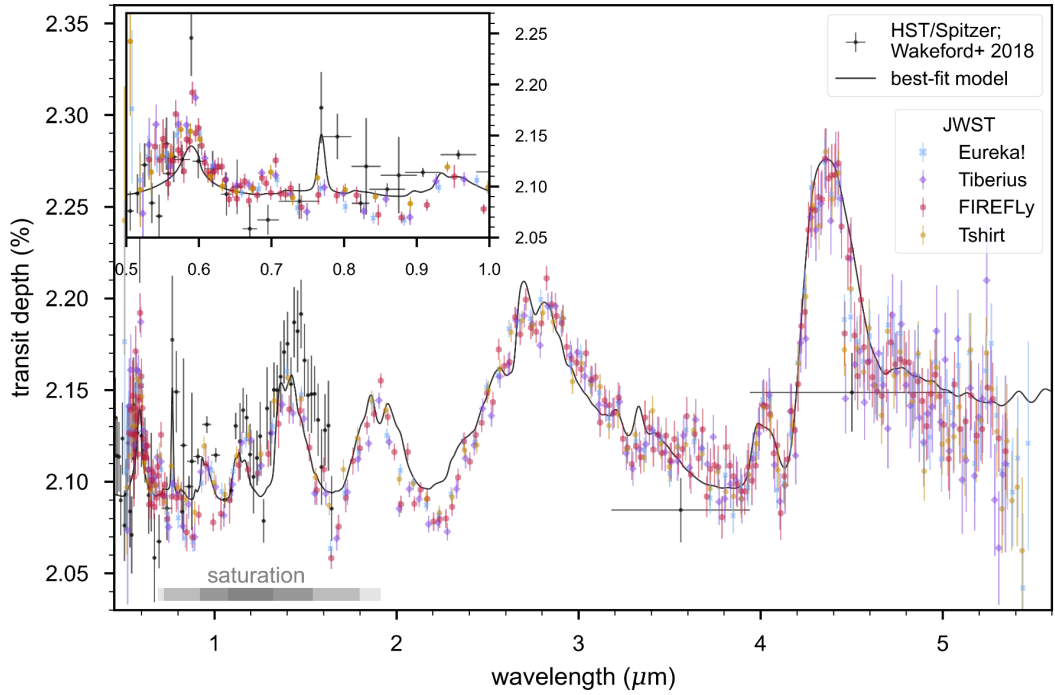

**Figure 3: WASP-39 b transmission spectral measurements**. A comparison of the JWST transmission spectra obtained from the four independent reductions considered in this work (coloured points), which are all in broad agreement. Previous measurements from HST, VLT, and Spitzer[6] are also shown (black) along with our fiducial best-fit spectrum model from the PICASO 3.0 grid (grey). All of the transmission spectral data have 1-$\sigma$ error bars shown. The saturated region of the detector is indicated (grey bar) with the shading representative of the level of saturation (also see Extended Figure 6). Different reductions are presented on slightly different wavelength grids for visual purposes, the original resolution each reduction used is discussed in the Methods.

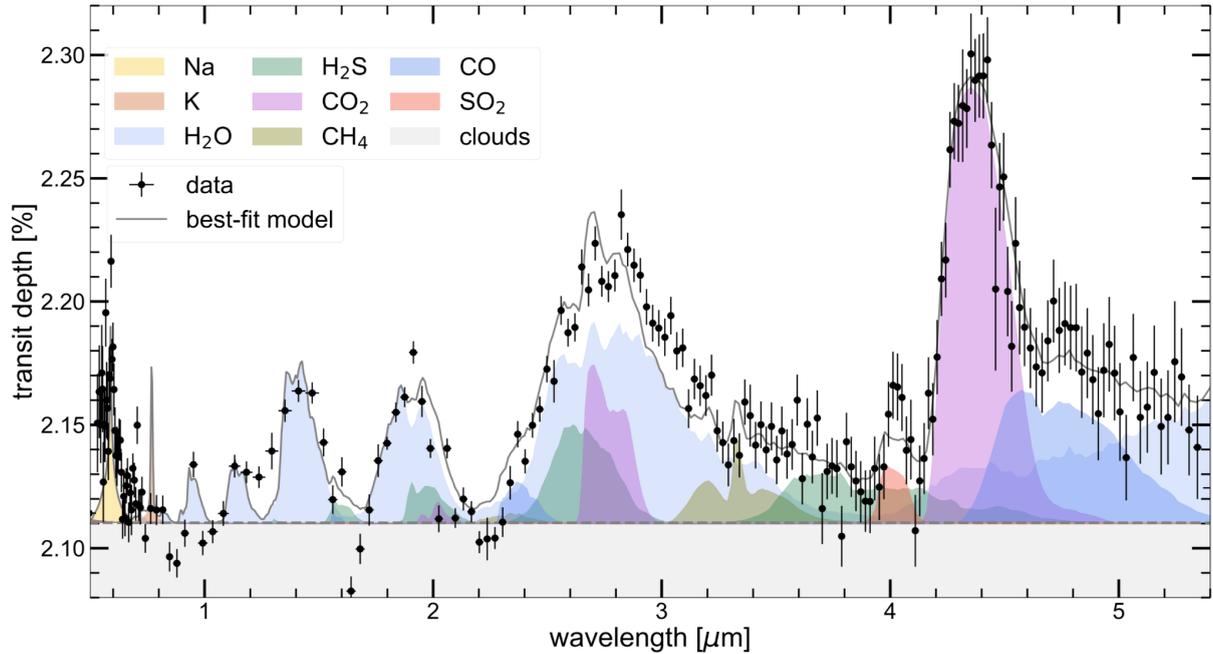

**Figure 4: The JWST-PRISM transmission spectrum of WASP-39b with key contributions to the atmospheric spectrum.** The black points with 1-$\sigma$ error bars correspond to the measured FIREFLy transit depths of the spectrophotometric light curves at different wavelengths. The best-fitting model spectrum from the PICASO 3.0 grid is shown as the grey line and the coloured regions correspond to the chemical opacity contributions at specific wavelengths. The best-fitting 1D radiative-convective thermodynamic equilibrium (RCTE) model corresponds to a super-solar metallicity and super-solar carbon-to-oxygen ratio with moderate cloud opacity (see Methods). The PRISM transmission spectrum is explained by contributions from Na (19$\sigma$), $H_2O$ (33$\sigma$), $CO_2$ (28$\sigma$), CO (7$\sigma$), $SO_2$ (2.7$\sigma$) and clouds (21$\sigma$). The data do not provide evidence of $CH_4$, $H_2S$ and K absorption (see Methods). Also, note that the detector was saturated to varying degrees between 0.8-1.9 $\mu$m.

# Methods

**Data Reduction**

One transit of WASP-39 b was observed with the NIRSpec PRISM mode, with the 8.23-hour observation roughly centred around the transit event. We used NIRSpec's Bright Object Time Series (BOTS) mode with the NRSRAPID readout pattern, the S1600A1 slit (1.6"×1.6"), and the SUB512 subarray. Throughout the exposure, we recorded 21,500 integrations, each with 5 0.28-second groups up the ramp. We achieved a duty cycle of 82%.

We extracted transmission spectra of WASP-39b using four different reductions with the FIREFLy, tshirt, Eureka!+ExoTEP, and Tiberius pipelines. The results from all reductions are broadly consistent (see Fig. 3 and Fig. 5). We used the FIREFLy reduction as our baseline for comparison to models throughout this paper, however equivalent overall results can be deduced from the other reductions. Some key attributes of the reductions are compared in Table 2. All reductions correct for $1/f$ noise: correlated frequency-dependent read noise in the images caused by detector readout and current biases in the electronics[31]. We note that since the GAINSCALE step of the JWST pipeline applies a gain correction to the raw count rate files, the counts and count rates quoted herein are in units of electrons and electrons per second, respectively.

We find that recovery of the saturated region was possible by applying several custom steps described here. Without these steps, the heavily saturated region showed a large and unexpected point-to-point scatter on the order of several thousand ppm in the transmission spectra. We note that there was limited on-sky NIRSpec calibration data available when the data were obtained and reduced, including an incomplete detector bias image whose values were all set to zero. We used a custom bias frame for this step (priv. comm., S. Birkmann). While the transmission spectra longward of about 2$\mu$m could be extracted without the use of this calibration, we found that bias correction was critical to extract the spectrum in the saturated region.

In addition, to recover the saturated region it was necessary to perform a reference pixel correction something that was skipped by the default jwst pipeline for NIRSpec PRISM because no official reference pixels are present in the sub array (also see t-shirt reduction below). All reductions also expand the saturation flags along entire columns and only use the groups prior to saturation for slope fitting in these regions. With these steps, the spectra broadly matched previous HST and VLT observations[6], with improvement in the region with only one or two groups before saturation. We expect that as updated NIRSpec calibration data becomes available the recovery of saturated regions in PRISM observations may become easier, however we still suggest avoiding rapid saturation with less than two groups prior to saturation if possible, especially if that region of the spectrum is important to one's science case.

    **FIREFLy**

We performed custom calibrations on the uncalibrated data, including 1/$f$ noise destriping[9] at the group level, bad and hot pixel cleaning, cosmic ray removal, and 5$\sigma$ outlier rejection. Destriping the data also removed potential background in the 2D images, though none was apparent in the data. The jump-step and dark-current stages of the jwst pipeline[12] (version 1.6.2) were skipped, and the top and bottom 6 pixels of the non-illuminated sub-array were manually set to be reference pixels in the jwst pipeline reference pixel step. To obtain our final wavelength calibration, we extrapolated the STScI-provided in-flight instrumental wavelength calibration data product across the detector edge pixels which did not have an assigned wavelength. The calibration was derived using the ground-based wavelength solution. We performed tests to search for zero-point offsets in the calibration versus the planetary and stellar spectra and did not find any at the level of half a pixel width or greater.

JWST detectors integrate using a non-destructive up-the-ramp sampling technique, where the flux is measured in counts-per-second from fitting the ramp from the groups contained within each integration. Fig. 6 shows the regions of the spectrum impacted by saturation. Within a column where a pixel was marked as saturated by the pipeline in any given group, we used only the data from the preceding groups for ramp fitting, and manually set an entire column of the detector as saturated if a pixel in that column was saturated. Because a small portion of the spectrum reaches our saturation threshold in the second group, this region of the spectrum only uses one group to derive a "ramp." While we were able to recover the spectra in this wavelength range by flagging and ignoring saturated pixels at the group level, we note that the data quality is lower in the saturated region than in the rest of the spectrum given the counts-per-second ramp was measured from fewer than the total 5 groups.

We measured the positional shift of the spectral trace across the detector throughout the time series using cross correlation and used them to shift-stabilize the images with flux-conserving interpolation. This procedure reduced the amplitude of position-dependent trends in the light curves. We optimized the width of our flux extraction aperture at each wavelength pixel and extracted the spectrophotometry. For each wavelength we tested a wide range of aperture widths and determined the width that minimized the scatter of the photometry of the first 350 datapoints. We bin the cleaned spectrophotometry in wavelength to create 207 variable-width spectral channels with roughly $10^5$ counts per second in each bin, and widths ranging from 3.3—60 nm. Because we use fewer groups in the saturated detector columns, our bin widths are larger by a factor of a few in this region to account for the lower count rates per detector column.

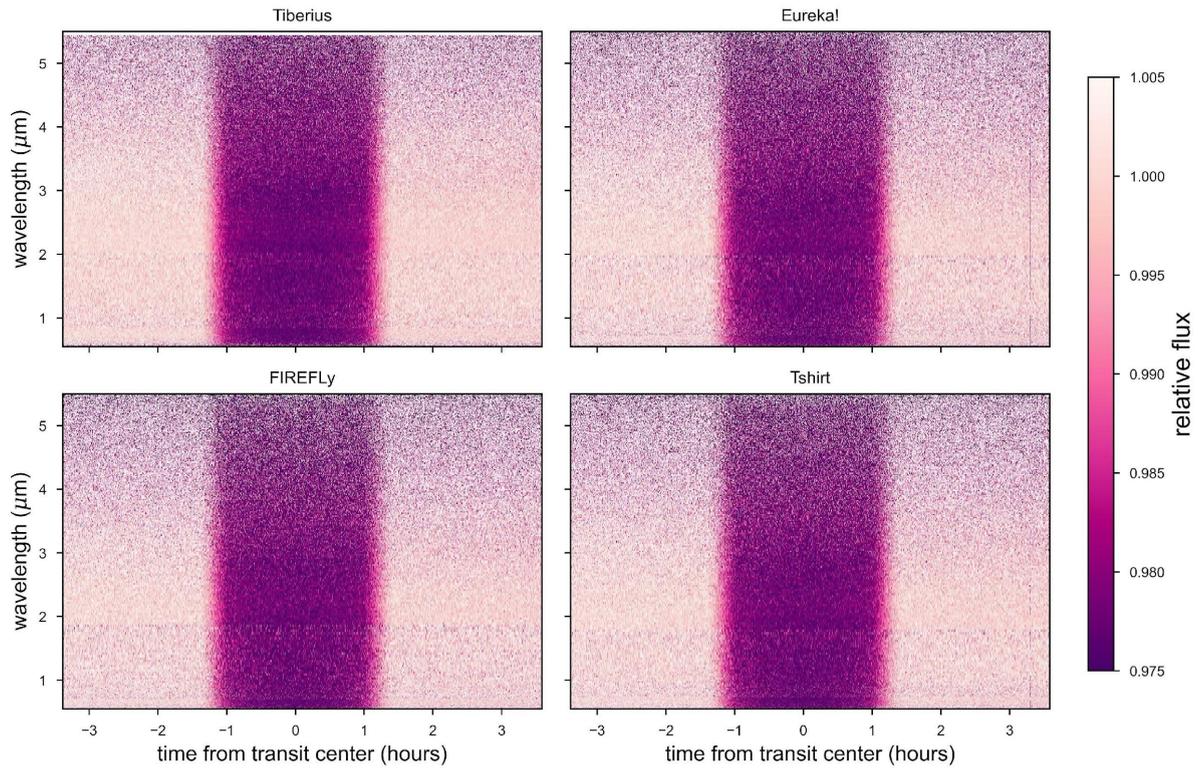

**Figure 5:** A comparison of the extracted 1D spectrophotometry across the four reductions.

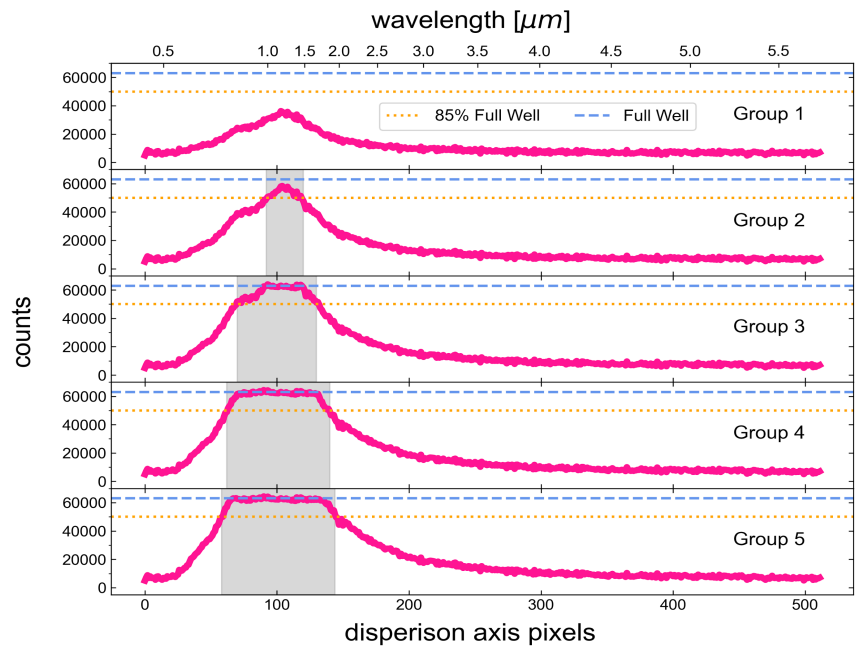

**Figure 6:** Demonstration of the impact of saturation. Shown are the group-level median frames from the uncalibrated data products across the entire integration. The dashed blue line represents the empirically derived saturation level, with the orange dotted line representing 85% saturation, the level adopted in the Eureka! reduction. Grey shaded regions represent columns that reach 85% full well in a given group.

**Table 1: Best-fit orbital parameters as measured from the FIREFLy white light curve.**

| Parameter | Value | Description |
|---|---|---|
| $T_0$ | 0.83532559 ± 0.0000007 | Mid-transit time [days] (BJD$_{TDB}$ - 2459770) |
| $a/R_s$ | 11.582±0.010 | Scaled semi-major axis |
| $b$ | 0.44537±0.00195 | Transit impact parameter |
| $\rho_s$ | 1.7873 ± 0.0048 | Stellar density [g cm$^{-3}$] (derived) |

Before fitting the transmission spectrum, we use a very wide, high-SNR white-light channel (3–5.5 µm) to fit for the planet's orbital parameters (listed in Table 1). Restricting the wide bin to the reddest wavelengths minimises the impact of limb darkening on the transit light curve and the resulting covariance with the orbital system parameters while ignoring the saturated region. We fit this white-light curve using the Markov Chain Monte Carlo sampler emcee[32] within the least-squares minimization framework of lmfit. We use 1,000 steps and uniform priors with extremely wide bounds that encapsulate the limits of physicality to ensure that there is no bias introduced by the prior. Our fitting approach accounts for non-Gaussian degeneracies in the posterior distribution, thereby addressing the known linear correlation between impact parameter ($b$) and the scaled semimajor axis ($a/R_s$).

We excluded the first 3000 integrations as they exhibited a slight non-linear baseline flux trend, and integrations 20750–20758 due to a high-gain antenna move which was identified from outliers in the photometry which correlated with noticeable trace shifts in the x- and y-directions. To measure the transmission spectrum, we fit the light curve at each wavelength channel jointly with a transit model[33] and a linear combination of systematics vectors composed of the measured spectral shifts in the x- and y-directions. At each channel we fit the planet's transit depth and the stellar limb darkening, while fixing the transit centre time $T_0$, impact parameter $b$, and normalised semimajor axis $a/R_s$ to the values determined in the white light curve fit. We also fix the orbital period to the published value of 4.0552941 days[34]. With the orbital system parameters fixed, we find the posterior distribution is well-fit by a multivariate Gaussian distribution, and therefore use a Levenberg-Marquart least squares minimization algorithm[25] to efficiently determine the best-fit parameters. In each channel, we inflate the transit depth error bars in quadrature with the measured residual red noise in the photometry as measured by the binning technique[35]. Measured uncertainties on the transit depths vary from 50–200 ppm, with a median of 99 ppm (see Fig. 8). As the noise levels are very close to the limit with what is expected including only photon and read noise sources, tools such as PandExo[36] should accurately predict what is achievable for other planets. We measure an increase in red noise for a few select spectral channels, but otherwise the light curves show no significant systematic errors, with some channels binning down to precision levels of a few ppm. We measure x− and y−jitter systematics at the ~100 ppm level. We see differences in the central transit time as a function of wavelength on the order of 10 seconds, which may be attributable to limb asymmetries in the atmospheric temperature and composition. We show these signatures in Fig. 7. Notably, we see a significant timing structure in the 2-3 μm range, which may arise from limb asymmetries in temperature and/or cloud coverage at the altitude probed by the water vapour absorption feature at 2.7 μm[37]. Further analysis of the spectrophotometry could be warranted to investigate limb asymmetries in more detail.

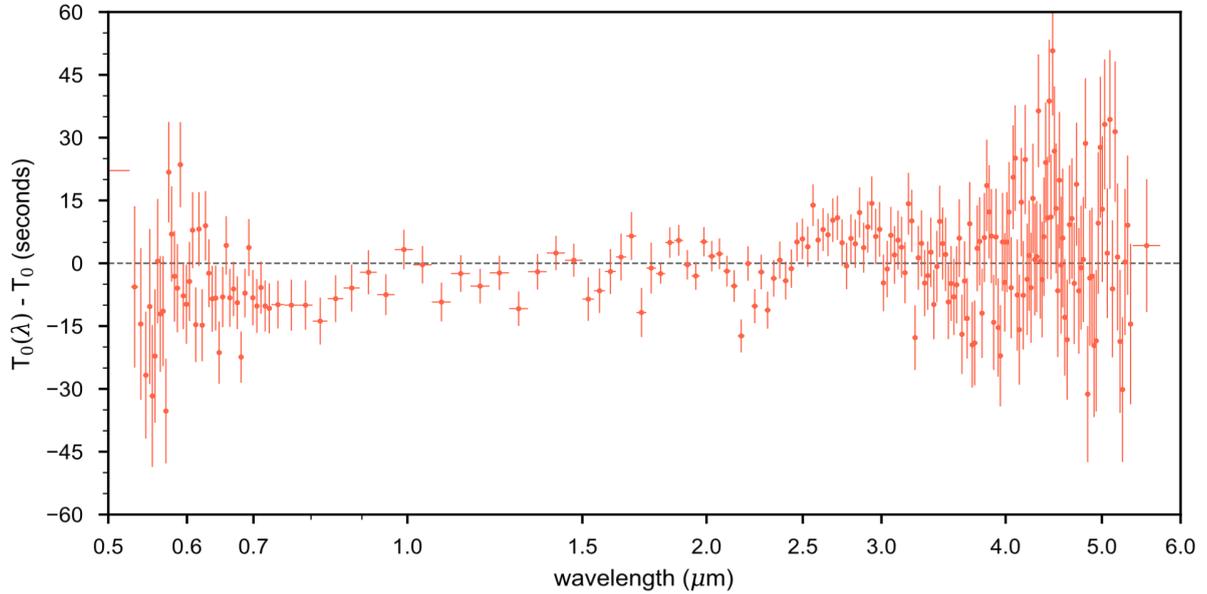

**Figure 7:** The wavelength-dependent central transit time in seconds. Structure is apparent–the prominent water and carbon dioxide absorption features at 2.7 *μ*m and 4.2 *μ*m, respectively, appear to arrive ~20 seconds after the optical continuum. A slope is also apparent from the blue side to the red.

We fit the transit light curves using a quadratic function to model stellar limb darkening given as,

$$\frac{I(\mu)}{I(1)} = 1 - a(1-\mu) - b(1-\mu)^2 \quad . \tag{1}$$

where $I(1)$ is the intensity at the centre of the stellar disk, $\mu = \cos(\theta)$ where $\theta$ is the angle between the line of sight and the emergent intensity, and *a* and *b* are the limb darkening coefficients. We tested a four-parameter non-linear limb darkening function[38] as well, which provided equivalent results. In practice, we first fit for both $u_+ = a + b$ and $u_- = a - b$ for the quadratic law. When comparing the limb darkening coefficients to theoretical values, we find an offset between the theoretically derived values of $u+$ from the 3D stellar models from ref.[39] and the JWST values derived from the transit light curve fits (see Fig. 9). This offset suggests the limb of WASP-39A is brighter than the stellar models predict. We fit for this offset and find it to be -0.065±0.022. As the wavelength-to-wavelength shape of $u_+$ is well described by the model, we then apply this offset to the theoretical limb darkening coefficients and then subsequently fix $u_+$ while allowing only $u_-$ to be free (see Fig. 9). This procedure helps reduce degeneracies when fitting multiple limb darkening coefficients and increases the precision of the transmission spectrum, as the limb darkening is often not well constrained, particularly at long wavelengths where the limb darkening is weak[39] (Fig. 9). The main effect of fitting for limb darkening over fixing the coefficients to the 3D models is the transit depth level of the optical spectrum, which is lower with values fixed to the model. We compare the optical spectrum with fixed limb darkening to the HST data from ref[6] in Fig. 10, which was also fit with limb darkening fixed to the same model. Overall, we find good agreement between the two spectra. We note that the assumptions around limb darkening can affect the optical spectra continuum which impacts particularly the interpreted levels of aerosol scattering — further investigations are warranted.

**tshirt**

We use the tshirt pipeline e.g.[41] to extract an independent set of light curves and spectrum. We begin with the uncalibrated "uncal" data product and apply a custom set of processing steps on stage 1 that build on the existing jwst stage 1 pipeline software version 1.6.0 with reference files CRDS jwst _0930.pmap. We use a custom bias file shared by the instrument team (Stephan Birkmann, private communication), which is the same file that was delivered to the JWST Calibration Reference Data System (CRDS).

We attempt to minimize the biasing effects of count rate non-linearity by modifying the quality flags of pixels surpassing 90% of full-well depth at the group stage. To ensure that there are no systematic differences between pixels within the spectral trace and in the background region, we adjust the quality flags uniformly along the entire pixel column at each group for all integrations. We skip the "jump" and "dark" steps of stage 1.

The tshirt code includes a Row-by-row, Odd-Even By Amplifier (ROEBA) correction to reduce $1/f$ noise. We first identify source pixels by choosing pixels with more than 5 Data Numbers per second (DN/s) in the rate file, and expanding this region out by 8 pixels. We then identify background pixels for $1/f$ corrections by choosing all non-source pixels and pipeline flagged non-'DO NOT USE' pixels. We loop through every group and subtract the median of odd (even) row background pixels from all odd (even) rows. We next find a column-by-column median of all background pixels to calculate a $1/f$ stripe correction and subtract this from each column.

After calculating rate files in DN/s, we use tshirt to perform covariance-weighted extraction of the spectrum[31]. We do a column-by-column linear background subtraction using pixels 0 through 7 and 25 through 32. We use a rectangular source extraction region centered on Y=16 pixels with a width of 14 pixels. We assume the correlation between pixels to be 8% from previous studies of background pixels[31]. We use a spline with 30 knots to estimate a smooth spectrum of the star at the source pixels and identify bad pixels as ones that deviate by more than $50\sigma$ from the spline. Pixels that are more than $50\sigma$ or else marked as 'DO NOT USE' are flagged and then the spatial profile is interpolated over those pixels. No corrections were made to the centroid or wavelength solution due to the exceptional pointing stability of the observatory[42].

When fitting the light curves, we exclude all time samples between UT 2022-07-10T23:20:01 and 202207-10T23:21:08 to avoid the effects of the high gain antenna move. We first fit the broadband light curve with all wavelengths. We assume zero eccentricity and the orbital parameters from[34] for $a/R_*$ and period. We try fitting the white light curve with eccentricity and argument of periastron set free and find that eccentricity is consistent with 0. We therefore assume zero eccentricity and a transit centre projected to the time of observations from a fit to the TESS data. We also assume an exponential temporal baseline in time to the data and a second-order polynomial trend in time. We fit the quadratic limb darkening parameters with uninformative priors[43] and the exoplanet code[44,45,46] with 3000 burn-in steps and 3000 sampling steps and 2 No U Turns Sampling chains[47]. We next binned the spectra into 116 bins, each 4 pixels wide. We fit all the individual spectroscopic channels with the orbital parameter fixed from the broadband light curve fit and only allowed the transit depth and limb darkening parameters to be free. Our resulting transit depth uncertainties ranged from 35 ppm to 732 ppm, with a median of 90 ppm.

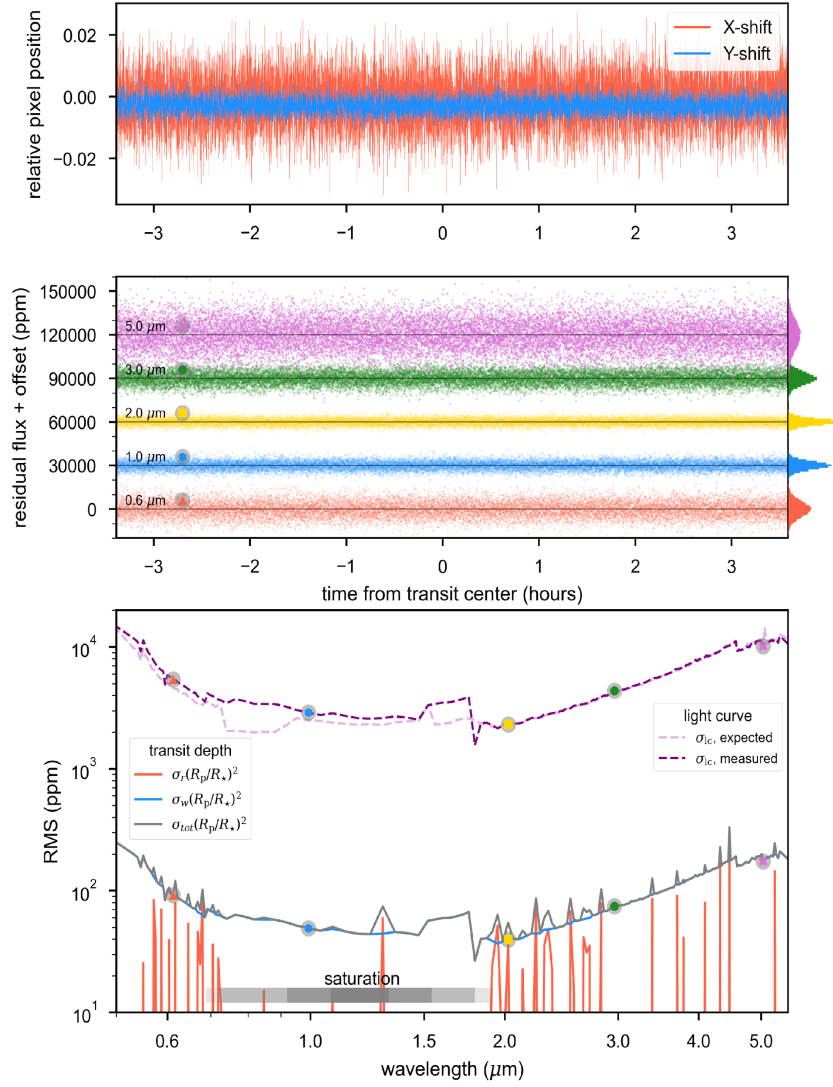

**Figure. 8: A summary of the positional shifts of the trace, the wavelength-dependent light curve scatter, and the transit depth noise.** (**Top**) The X- and Y-shift vectors as measured by 1D cross correlation with FIREFLy. (**Middle**) The residual spectrophotometric light curves are shown for four representative spectral channels spanning the PRISM wavelength range with no temporal binning. The residual scatter is approximately Gaussian for each, as indicated by the histogram on the right y-axis. We validate this by performing Anderson-Darling tests on the residuals of the spectral and white-light curves, and find that all of the Anderson-Darling test statistics lie below the respective critical values 1% significance level. Therefore, we find that there is not sufficient evidence that the residuals are not normally distributed. (**Bottom**) The top two purple curves show the expected and measured normalised light curve root mean square (RMS) residuals, with no temporal binning. Longward of 2 $\mu$m, the scatter in each light curve matches well with the expected noise as estimated by the jwst pipeline, which is dominated by photon noise. This agreement indicates the majority of the light curves reach near the photon limit. The transit depth uncertainties are also plotted below, including the white noise (blue, $\sigma_w$), red noise (red, $\sigma_{red}$), and total noise components (grey, $\sigma_{tot}$). Some wavelength bins have enhanced red noise, but the majority of the transmission spectrum is consistent with minimal red noise from residual

systematic errors. The wavelengths affected by detector saturation are indicated by the grey shaded bar, with darker colors corresponding to quicker saturation. The colored dots are the measured RMS values from the light curves shown in the top panel.

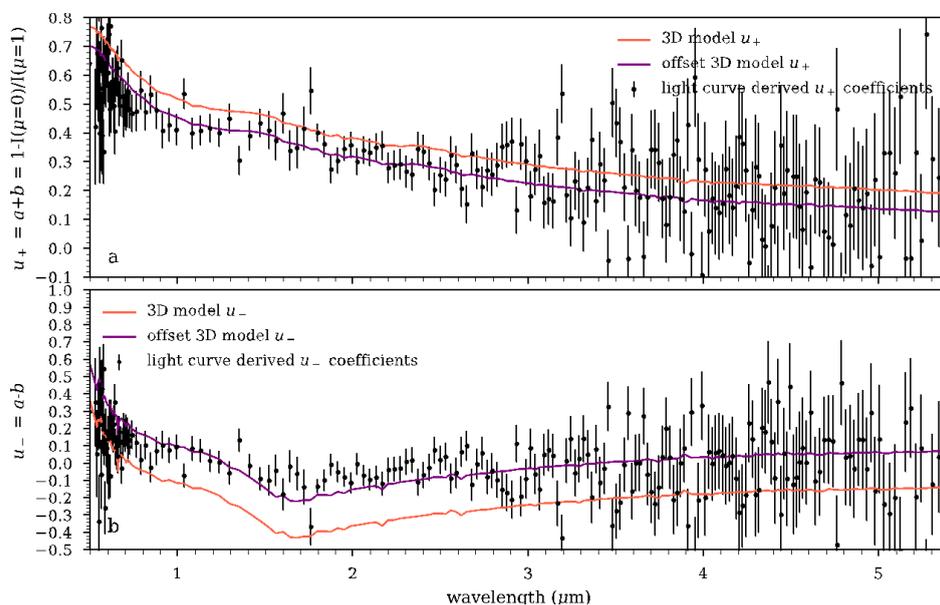

**Figure 9: Empirically derived stellar limb darkening coefficients fit with a quadratic law. a,** the fit $u_+$ coefficients (black) along with the theoretically predicted values derived from a 3D stellar model (red). The theoretical $u_+$ values with a constant offset of -0.065±0.022 (purple) is also shown. The theoretical models predict the wavelength-to-wavelength shape of $u_+$ well. As $u_+$ is directly related to the intensity of the star at the stellar limb ref.[40], these findings suggest WASP-39A is 6% brighter at the limb than models predict. **b**, similar as **a**, but for the $u_-$ coefficient. As the shape of the derived coefficients differs from the model prediction, $u_-$ was left free to vary in the transmission spectral fits.

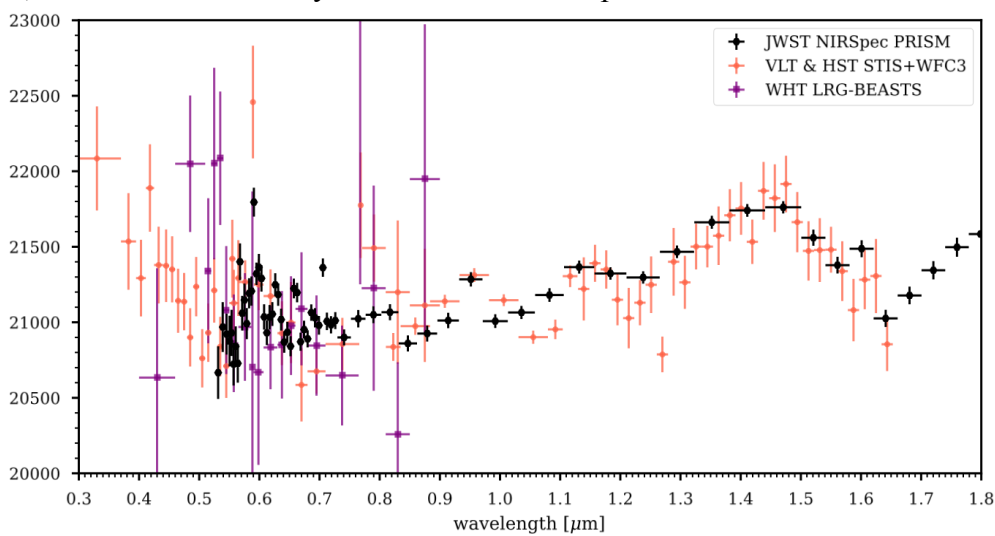

**Figure 10:** Comparison of the JWST NIRSpec PRISM data (black) for WASP-39b along with HST and VLT data from ref.[15,6] and WHT data from ref.[16]. The JWST spectrum was derived with the limb darkening fixed to the same 3D stellar model as in[6] to aid comparisons. With fixed limb darkening, the JWST transmission spectrum has lower overall transit depths especially at optical wavelengths. The broadband spectrum from the two space telescopes

compares well, including the amplitude of the 1.4$\mu m$ water feature first observed by HST/WFC3 and the Na feature near 0.6$\mu m$ observed by HST/STIS.

**Eureka! and ExoTEP**

We use the Eureka! pipeline[48] for the data reduction steps of detector processing, data calibration and stellar spectrum extraction, and the ExoTEP pipeline[49,50,51] to generate light curves in each wavelength bin and perform light curve fitting.

We start our data reduction using the uncalibrated "uncal" outputs of the jwst pipeline's Stage 0. From there, Eureka! acts as a wrapper for the first two stages of the jwst pipeline, version 1.6.0. We use the jwst pipeline to fit slopes to the ramp in each pixel and perform data calibration, and follow the default pipeline steps unless otherwise stated. We skip the jump detection step, meant to correct the ramps for discontinuities in the slopes of group count rates as a function of time. Due to the small number of groups up the ramp, performing this step leads to a large fraction of the detector pixels being incorrectly flagged as outliers, and we therefore rely on the time series outlier clipping steps in the subsequent stages to correct for cosmic rays. A custom bias frame is used, rather than the default one available on CRDS at the time of reduction. We also expand the saturation flags in Stage 1 to ignore saturated pixels more conservatively than allowed by the default jwst pipeline settings: in each group, we flag pixels as saturated if they reach ~85% of the full well in the median image across all integrations for that group and expand the saturation flag such that in a given detector column (constant wavelength) all pixels are marked as saturated if any one pixel in that column is flagged. This is implemented by inputting the indices of columns to mask based on inspection of the uncal data products, rather than an internal calculation of the full well percentage. We include a version of the ROEBA correction described above, using the top and bottom 6 rows. We further add a custom background correction at the group level prior to ramp fitting, and subtract from each column the median of the six pixels at the top and at the bottom of the detector, excluding outliers at more than the 3-$\sigma$ level. We skip the "photom" step in Stage 2 of the STScI detector pipeline because absolute fluxes are not needed in our analysis. We also skip the "extract1d" step as we perform custom spectral extraction using Eureka!.

For 1D spectral extraction, we trim the array to include only columns 14 to 495 in the dispersion direction, as NIRSpec's throughput is negligible beyond this range. We then use the median detector frame to construct the weights used in the optimal extraction based on[52]. Pixels are masked if they have an marked data quality flag (i.e., bad pixels that are flagged by the jwst pipeline as "DO NOT USE" for various reasons) or if they are clipped by two iterations of 10-$\sigma$-clipping of the time series. We perform the optimal extraction over 8 rows centered on the source position (corresponding to a spectral half-width aperture of 4 pixels). The source position is identified from the maximum of a Gaussian fitted to the summed spatial profile from all detector columns over the entire integration.

We use ExoTEP to generate median-normalized light curves at the native pixel resolution from each detector column, using the stellar spectra outputs from Stage 3 of Eureka!. We then perform further clipping of outliers in time in the white and wavelength-dependent light curves by computing a running median with a window size of 20 and excluding 3$\sigma$ outliers in several time series. This outlier-clipping was applied to the flux, source position and width in the cross-dispersion direction in each frame and spectrum shifts in the dispersion direction.

We jointly fit astrophysical and systematics model parameters to the white (0.5–5.5$\mu$m) light curves and each of the wavelength-dependent light curves. Our astrophysical transit model is calculated using the batman package[33]. Using the white light curve, we fit for the two coefficients of a quadratic limb darkening law (Equation 1), WASP-39b's impact parameter, scaled semi-major axis $a/R_s$, time of transit centre, and the planet-to-star radius ratio. In each of the wavelength channels we then fix the planet's impact parameter, semi-major axis and transit time to the values derived from the white light curve and fit only for the planet-to-star radius ratio and the two quadratic limb darkening coefficients. For the systematics model, we assume a linear trend with time that can be different in each spectroscopic channel, and fit for its slope and y-intercept. Lastly, we fit a single-point scatter to each light curve, which illustrates the level of scatter required for our joint model to reach a reduced chi-squared of 1. The fitted light curve scatter in both the white light curve and wavelength-dependent channels is within a few percent of the expectation from the high-frequency scatter in the raw light curves, which attests to the lack of systematics. We bin the final transmission spectrum (four points binned together throughout the spectrum) for visual comparison with the other reductions in Figure 3.

**Tiberius**

The Tiberius pipeline builds upon the LRG-BEASTS spectral reduction and analysis pipelines introduced in[53,16,54]. The Tiberius pipeline operates on the Stage 1 JWST data products to obtain 1D stellar spectra via tracing of the stellar spectra, fitting and removal of the background noise, and simple aperture photometry. We used the FIREFLy-processed Stage 0 data.

Prior to tracing the spectra, we interpolate each column of the detector onto a finer grid, 10× the initial spatial resolution. This step improves the extraction of flux at the sub-pixel level, particularly where the edges of the photometric aperture bisect a pixel, and leads to a 14% reduction in the noise in the data. We also interpolate over the bad pixels using their nearest neighboring pixels in $x$ and $y$. We identify bad pixels by combining 5$\sigma$ outlying pixels found via running medians operating along the pixel rows with bad pixels identified by visual inspection. We trace the spectrum by fitting a Gaussian distribution at each column (where a column refers to the cross-dispersion direction) to the stellar spectra. We then use a running median, calculated with a moving box with a width of five data points, to smooth the measured centres of the trace. We fit these smoothed centres with a fourth-order polynomial, removed five median absolute deviation outliers, and refitted with a fourth-order polynomial.

To remove residual background flux not captured by the 1/$f$ correction, we fit a linear polynomial along each column in the spatial direction. We mask the stellar spectrum, defined by an aperture with a full width of 4 pixels centered on the trace we found in the previous step, from this background fit. We also mask an additional 7 pixels on either side of this aperture so that the background fit is not impacted by the wings of the stellar PSF. This left us with 7 pixels at each edge of the detector (a total of 14 pixels) to estimate the background with. We also clipped any pixels within the background that deviate by more than three standard deviations from the mean for that particular column and frame to avoid residual bad pixels and cosmic rays impacting our background estimation. We found that this additional background step led to a 3% improvement in the precision of the transmission spectrum.

The stellar spectra are then extracted by summing the flux within a 4-pixel-wide aperture following the removal of the background at each column. The background count level, as estimated by the JWST Exposure Time Calculator (ETC) is on the order of a few counts per second, meaning the background is negligible. Further, since we perform 1/f subtraction, this faint background is subtracted column-by-column. We experimented with the choice of the aperture width, also running reductions with 8- and 16-pixel-wide apertures. The 8-pixel-wide aperture gave a median uncertainty 1% larger than a 4-pixel aperture and a 16 pixel aperture gave an uncertainty 15% larger than 4-pixels. This same change was reflected in the median RMS of the residuals to the light curve fits. Since the stellar PSF is so narrow in PRISM data, we believe that the increase in noise with increasing aperture width is related to the increasing influence of photon noise, readnoise and bad pixels where the stellar flux is lower. Following the extraction of the stellar spectra, we divide the measured count rates by a factor of 10 to correct for our pixel oversampling, as described above.

To remove residual cosmic rays, we identify outliers in each stellar spectrum via comparison with the median stellar spectrum. We did this in three iterations, each of which involves making a median spectrum, identifying outliers (10, 9, 8 $\sigma$) and replacing pixels containing a cosmic ray with a linear interpolation between neighboring pixels. We tested this interpolation against assigning the cosmic ray pixels zero weight and found that this led to a negligible difference in the transmission spectrum. To correct for shifts in the stellar spectra and align each spectrum in pixel space, we cross-correlate each stellar spectrum with the first spectrum of the observation and linearly resample each spectrum onto a common wavelength grid. We adopt the custom wavelength solution calculated by the tshirt pipeline, which uses the jwst pipeline to evaluate the wavelengths at pixel row 16 using the world coordinate system.

Our white light curves are created by summing over the full wavelength range between 0.518–5.348$\mu$m. We make two sets of spectroscopic light curves: one set of 440 light curves at 1-pixel resolution and one set of 147 light curves at 3-pixel resolution. We mask integrations 20751–20765 due to a high gain antenna move that leads to increased noise in the light curves. We also mask the first 2000 integrations from our analysis due to a systematic ramp. This means our light curves each contained 19486 data points.

To fit our light curves, we began by fitting the white light curve to determine the system parameters.

We fit for the following parameters: the scaled planetary radius ($R_p/R_s$), the planet's orbital inclination ($i$), the time of mid-transit ($T_C$), the scaled separation ($a/R_s$), the linear limb darkening coefficient ($u_1$), and the parameters defining the systematics model. We fix the planet's orbital period to 4.0552941d and eccentricity to 0[34]. For the remaining parameters, we use the values from[34] as initial guesses.

For the analytic transit light curve model, we use batman[33] with a quadratic limb darkening law. We use ExoTiC-LD[55,56], with 3D stellar models[39] to determine the appropriate coefficients, adopting the stellar parameters ($T_{eff}$ = 5512±55K, log$g$ = 4.47±0.03 cgs, [Fe/H] = 0.01±0.09 dex) from[34] and Gaia DR3[57,58]. For our final fits, we fix the quadratic coefficient, $u_2$, to the values determined by ExoTiC-LD. However, we also run a set of fits with neither $u_1$ nor $u_2$ fixed and find this leads to a transmission spectrum that is qualitatively similar to the one in which LDs are fixed. For the systematics model, we sum the following three polynomials: quadratic in time, linear in $x$ position of the star on the detector, and linear in $y$ position of the star on the detector. The final fit model, $M$, was of the form:

$$M(t) = T(t,p) \times (\Sigma_i(S_i(a_i,s)^{n_i})) \qquad (2)$$

Where $t$ is time, $p$ are the parameters of the transit model, $T$, $a$ are the ancillary data, and $s$ are the parameters (polynomial coefficients) of the systematics model, $S$. The systematics model is the sum of the polynomials operating over each ancillary input, $a_i$, with $n_i$ defining the order of the polynomial used for each input.

We fit our white light curve in three steps: a first fit to remove any $4\sigma$ outliers from the light curves, a second fit that is used to rescale the photometric uncertainties such that the best-fitting model gives $\chi_v^2 = 1$, and a third fit with the rescaled photometric uncertainties, from which our final parameter values and uncertainties are estimated. The parameter uncertainties were calculated as the standard deviation of the diagonal of the covariance matrix that was in turn calculated from the Jacobian returned by scipy.optimize.

Following the white light curve, we fit our spectroscopic, wavelength-binned, light curves. For these fits, we held $a/R_s$, $i$, and $T_C$ fixed to the values determined from the white light curve fit: 11.462 ± 0.014, 87.847 ± 0.015 deg, 2459770.835623 ± 0.000008 BJD$_{TDB}$. These values are somewhat different to the FIREFLy reduced white light parameters, and these differences will be explored in greater detail in a future work. To zeroth order, offsets in orbital parameters result in simple vertical offsets in the resulting transmission spectrum. The remaining fit parameters were the same as for the white light curve fit. We perform the same iteration of fits using a Levenberg–Marquardt algorithm to determine $R_p/R_s$ as a function of wavelength.

**Reduction Comparison**

Procedural differences exist across the four main reductions of the dataset, which may account for the subtle qualitative differences between the final reduced spectra. A careful investigation of these nuances is warranted and will be presented in a future paper. Table 2 highlights some key procedural differences between the reductions. We note that despite these differences, the resulting exoplanet spectra are qualitatively in excellent agreement with each other (see Fig. 3), owing to the stability of the data and the self-calibrating nature of the transit technique.

| Reduction step | FIREFLy | Tshirt | Eureka! | Tiberius |
|---|---|---|---|---|
| Background, $1/f$ subtraction | y | y | y | y |
| X-, Y-shift correction | y | n | y | y |
| X-, Y-shift detrending | y | n | y | y |
| Baseline detrending | y | y | y | y |
| Trace extraction optimization | y | y | y | n |
| Pre-transit baseline trim | y | y | n | y |
| **Mean spectrophotometric scatter (ppm)** | 676 | 725 | 815 | 709 |

**Table 2**: An overview of the analysis procedures used by the independent data reductions. The spectrophotometric scatter is estimated from the standard deviation of the pre-transit data between 0.62-5.42 $\mu$m with only a linear baseline trend removed

**Stellar Activity**

WASP-39b has a reported low activity level[8] , with a Ca II H and K stellar activity index of logR'$_{HK}$=-4.994[ref. 4]. NGTS and TESS photometric monitoring of WASP-39A is reported in ref.23, which finds low modulations at the 0.06% level with no apparent star-spot crossings. With low stellar activity levels, the transit observations are unlikely to be affected by stellar activity.

**Forward Model Grids**

We use four different 1D radiative–convective–thermochemical–equilibrium (RCTE) model grids to assess atmospheric properties like detection of individual gases, metallicity, carbon–to–oxygen (C/O) elemental abundance ratio, and the presence/absence of clouds. The ScCHIMERA[59, 60], PICASO 3.0[61,62,63,64], ATMO[65,55,67], and PHOENIX[68,69] models were used to generate these grids specifically for WASP-39 b. While the ATMO and the PHOENIX grids were used to fit the data with a reduced $\chi^2$ based grid search method, the PICASO 3.0 and ScCHIMERA grids were used in a grid retrieval framework using a nested sampler[70,71] . Within each nested sample likelihood calculation, the transmission spectra are generated on-the-fly by post-processing the pre-computed 1D RCTE model atmospheres. The $SO_2$ volume mixing ratio and cloud properties are injected into spectrum during this post-processed transmission calculation. Fig. 11 shows best-fit models obtained by each of the four grids compared with the transmission spectrum obtained with the FIREFLy data reduction pipeline. ScCHIMERA, PICASO 3.0, and ATMO produce fits with reduced $\chi^2$ between 3.2–3.3, while the PHOENIX grid obtains a reduced $\chi^2$ of 4.3. The reduced $\chi^2$ is defined as the total $\chi^2$ calculated from all the data points divided by the total number of data points. While PICASO 3.0, ScCHIMERA, and ATMO predict the metallicity of the atmosphere to be about 10×solar, PHOENIX finds a best-fit metallicity to be a 100×solar which might be due to the larger grid spacing of the PHOENIX grid along both the cloud and metallicity dimensions. While the models qualitatively match the data, the reduced $\chi^2$ obtained by the best-fitting models from these grids are also > 3, which suggests that these are not fitting the data particularly well. These relatively poor fits could arise for multiple reasons, such as the region of the data affected by saturation, the presence of disequilibrium chemistry in the atmosphere due to vertical mixing or photochemistry, and the non-grey nature of scattering in the upper atmosphere. Table 3 provides a summary of the best-fit atmospheric parameters obtained by the four different grids with different fitting methods (grid retrievals and grid search). In order to explore the effect of the saturated region on the best-fit parameters, we inflate the transit depth errors in the saturated regions (0.68 $\mu$m – 1.91 $\mu$m) by a factor of 1000 and recompute the best-fit models using the grid retrieval framework with both the PICASO 3.0 and ScCHIMERA grids. We find that this did not significantly change any of the best-fit parameters including the metallicity and the C/O ratio. Table 3 lists the best-fit parameters obtained when the saturated region error bars were inflated by a factor of 1000.We summarize the main results obtained by these 1D grids here and refer the reader to ref. 22 for detailed descriptions of each of these model grids.

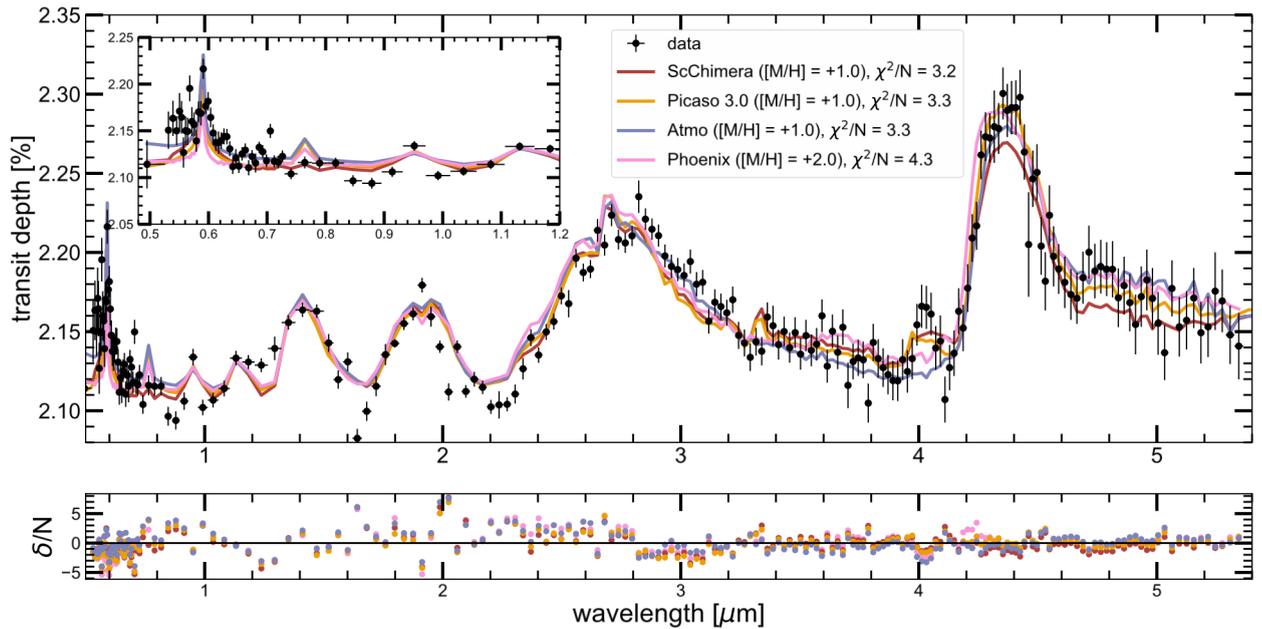

**Figure 11:** Best-fit models from ScCHIMERA, PICASO 3.0, ATMO, and Phoenix 1D RCTE model grids for WASP-39b, with the FIREFLy reduction overlaid are shown in the top panel. The top left inset panel shows the data and the models between 0.5-1.2 $\mu$m. All these models prefer super-solar atmospheric metallicities and cloudy atmospheres for WASP-39 b. The C/O ratio estimated by these models lies in the range 0.6– 0.7. Additional $SO_2$ was injected in the PICASO 3.0 and ScCHIMERA grids to estimate the abundance of $SO_2$ required to explain the 4.0 $\mu$m feature, in a Bayesian framework. The ATMO and PHOENIX models are shown without any additionally injected $SO_2$ to emphasize that RCTE models do not predict such an $SO_2$ feature and chemical disequilibrium effects are required to explain the observed feature. The bottom panel shows the residuals from each best-fit model divided by the noise in the transit depth as a function of wavelength.

**Table** 3: Overview of the best-fit model parameters obtained from each grid. PICASO 3.0 and ScCHIMERA grids follow the grid retrieval (GR) framework to obtain the best-fit models whereas ATMO and PHOENIX use the reduced $\chi^2$ minimization based grid search method (GS). To test the effect of the saturated region on the obtained best-fit parameters, the PICASO 3.0 and ScCHIMERA grid were used to also do a fit with the error bars in the saturated region (0.68 $\mu$m – 1.91 $\mu$m) inflated 1000 times. The best-fit parameters did not show any significant change due to this exercise but are still listed in the table under the w/o SR column. The best-fit parameters obtained by fitting the full spectrum are listed under the w/ SR column. Note that even though the w/o SR fits were obtained by inflating the errorbars in the saturated region, the reduced $\chi^2$ reported in the w/o SR column are computed without the points in the saturated region for direct comparison with the reduced $\chi^2$ obtained from fitting the full spectrum. Also, note that the ATMO models include cloud opacities with an adjustable multiple of the $H_2$ Rayleigh scattering opacity at 350 nm. Therefore the 5×$H_2$ in this table for the ATMO grid corresponds to a gray cloud opacity which is 5× the $H_2$ Rayleigh scattering opacity at 350 nm between 1 to 50 mbar pressures

| Parameter | PICASO 3.0 w/ SR | PICASO 3.0 w/o SR | ScCHIMERA w/ SR | ScCHIMERA w/o SR | ATMO w/ SR | PHOENIX w/ SR |
|---|---|---|---|---|---|---|
| [M/H] | +1.0 | +1.0 | +1.0 | +1.0 | +1.0 | +2.0 |
| C/O | 0.68 | 0.68 | 0.65 | 0.65 | 0.7 | 0.9 |
| $\kappa_{cld}$ [cm$^2$/g] | $10^{-2.07}$ | $10^{-2.04}$ | $10^{-2.46}$ | $10^{-2.52}$ | 5×$H_2$ | Opaque |
| $P_{cld}$ | – | – | – | – | 1-50 mbar | 1 mbar |
| Rayleigh Scattering | $H_2$ only | $H_2$ only | $H_2$ only | $H_2$ only | 10×multigas | $H_2$ only |
| $\log_{10}(SO_2)$ | -5.2 | -5.1 | -5.7 | -5.7 | – | – |
| $\chi^2$/N | 3.3 | 3.2 | 3.2 | 2.9 | 3.3 | 4.3 |
| Method | GR | GR | GR | GR | GS | GS |

**Detection Significance of Gases**

We quantify the detection significance of each species through a Bayes factor analysis[e.g.,72]. To do so within the ScCHIMERA grid retrieval framework, we remove each gas during the transmission spectrum computation step (the 1D RCTE atmosphere models remain unchanged) one at a time and re-run the nested sampler. We compare the Bayesian evidence of each removed-gas run to that of the grid retrieval with all of the gases. There is no change in the number of parameters with the exception of the cloud and $SO_2$ mixing ratio parameters. Table 4 shows the result of this exercise summarized as the log-Bayes factor and a conversion to the detection significance[e.g.,73].

We also quantify the detection significances of different gases following the procedure used in ref. [22]. To calculate the detection significance of each gas, the best-fit transmission spectrum model from the PICASO 3.0 grid ([M/H] = +1.0, C/O= 0.68) is re-calculated without that gas. The wavelength ranges where the particular gas has the most prominent effect are first identified and then a residual spectrum is calculated by subtracting the model without the gas from the data. The residual spectra for $H_2O$, $CO_2$, CO, Na, $SO_2$ and $CH_4$ are shown in the six panels of Fig. 12. We fit each of these residual spectra with two functions, a Gaussian/double Gaussian/Voigt function and a constant line. We use the Dynesty nested-sampling routine to perform the fits and to determine the Bayesian evidence associated with each fit. The Bayes factor between the fits of the residual spectrum with the Gaussian/Voigt function and the constant line is then used to determine the detection significance of a gas. For example, for computing the detection significance of $H_2O$, two adjacent $H_2O$ features between 1 and 2.2 $\mu$m are used. We note that $H_2O$ is expected to be the dominant opacity source in other wavelength ranges (e.g., 2.2–3 $\mu$m) as well, so choosing two features for this analysis would produce a lower limit on the detection significance of $H_2O$. The best–fit double Gaussian function to these features along with its 1$\sigma$ and 2$\sigma$ envelopes are shown with the red line and shaded regions in Fig. 12 top–left panel. The same residual spectrum is also fitted with a straight line shown with blue colour in Fig. 12. The logarithm of the Bayes factor between the two models is found to be ln$B$=242, which shows that the model with $H_2O$ is significantly favored over a model without any $H_2O$. The detection significance of $H_2O$ corresponding to this Bayes factor is calculated using the prescription in ref. 73 and is found to be 22$\sigma$. The same methodology, but with a single Gaussian function, is also followed for $CO_2$, CO, $SO_2$, $H_2S$, and $CH_4$ to get their detection significance summarized in Table 4 last column. Our Gaussian residual fit significance for $CO_2$ matches the initial analysis of the NIRSpec PRISM data presented in ref.[22].

As shown in Table 4, the detection significance of all gases increases with the Bayes factor analysis technique relative to the Gaussian/Voigt function technique. This is notably also the case for $SO_2$, lending confidence to the detection and identification of the molecule, as the feature is better fit by its respective opacity profile.

**Resolution Bias and the detection Significance of CO**

The Resolution-Linked Bias effect (RLB) serves to dilute the measured amplitudes of planetary atmospheric features due to overlapping absorption lines in the stellar atmosphere. While this effect is negligible for most stars earlier than M dwarfs, some stellar CO absorption is expected in WASP-39, meaning the measured planetary CO abundance may be biased. Following Eq. 4 of ref.[74] and using high-resolution (R ~ $10^5$) PHOENIX models of the planet and the star, we quantify an upper limit on the magnitude of this bias effect. We find that the planetary CO feature is biased by 30 to 40 ppm in the 4.5-5.1 $\mu$m region, leading to as much as a ~1-$\sigma$ underestimate of the planetary CO absorption strength, and subsequently a similar underestimate of its abundance. We note that this effect is potentially weakened by Doppler broadening of the molecular lines (which is unaccounted for by PHOENIX) due to stellar rotation, planetary orbital radial velocity, and planetary winds. Future work, which may benefit from more detailed modeling and high-resolution observations of WASP-39's CO band heads, will better quantify the magnitude of this dilution.

**Metallicity, C/O Ratio and $CH_4$ abundance**

The best–fitting atmospheric metallicity for WASP-39 b is found to be ~10× the solar metallicity using the model grids. The top panel in Fig. 13 shows the observed transmission spectrum of the planet between 2.0–5.3 $\mu$m (where variations due to metallicity are most prominent), along with multiple transmission spectrum models assuming different atmospheric metallicities ranging from sub-solar values (e.g., 0.3×solar) to super-solar values (e.g., 100×solar). The bottom panel demonstrates the effect of different atmospheric C/O ratios at 10×solar metallicity on multiple transmission spectrum models along with the data. Since the star WASP-39 has near-solar elemental abundances[83], scaled solar abundances are a reasonable choice for this star. The $CH_4$ feature between 3.1–4 $\mu$m and 2.2–2.5 $\mu$m is very prominent in sub-solar and solar metallicity thermochemical equilibrium models shown in Fig. 13. The absence of such a $CH_4$ feature in the data is evident. This, combined with the large $CO_2$ feature between 4.3–4.6 $\mu$m and measurable CO feature at 4.7 $\mu$m, led to a super-solar (10×) metallicity estimate for the planet. The C/O ratio of the RCTE models significantly affects the predicted gas abundances, and therefore the calculated transmission spectrum. Fig. 13 bottom panel shows that for metal-rich atmospheres (e.g., >10× solar) with C/O ratios lower than 0.7, the transmission spectrum is dominated by features of oxygen-bearing gases ($H_2O$, $CO_2$, CO) [e.g., 84,85,67]. But for higher C/O ratios (e.g., 0.916), the transmission spectrum becomes $CH_4$ dominated at wavelengths greater than 1.5 $\mu$m. We obtain an upper limit on the C/O ratio of WASP-39 b at about ~ 0.7. However, these interpretations are based on single-best fits from model grids assuming thermochemical equilibrium. Other chemical disequilibrium processes like atmospheric mixing and high–energy stellar radiation-induced photochemistry can also potentially affect this interpretation. These disequilibrium chemistry effects require further exploration in the context of WASP-39 b and will be discussed in future work (Welbanks et al. (in prep), Tsai et al. (in prep)).

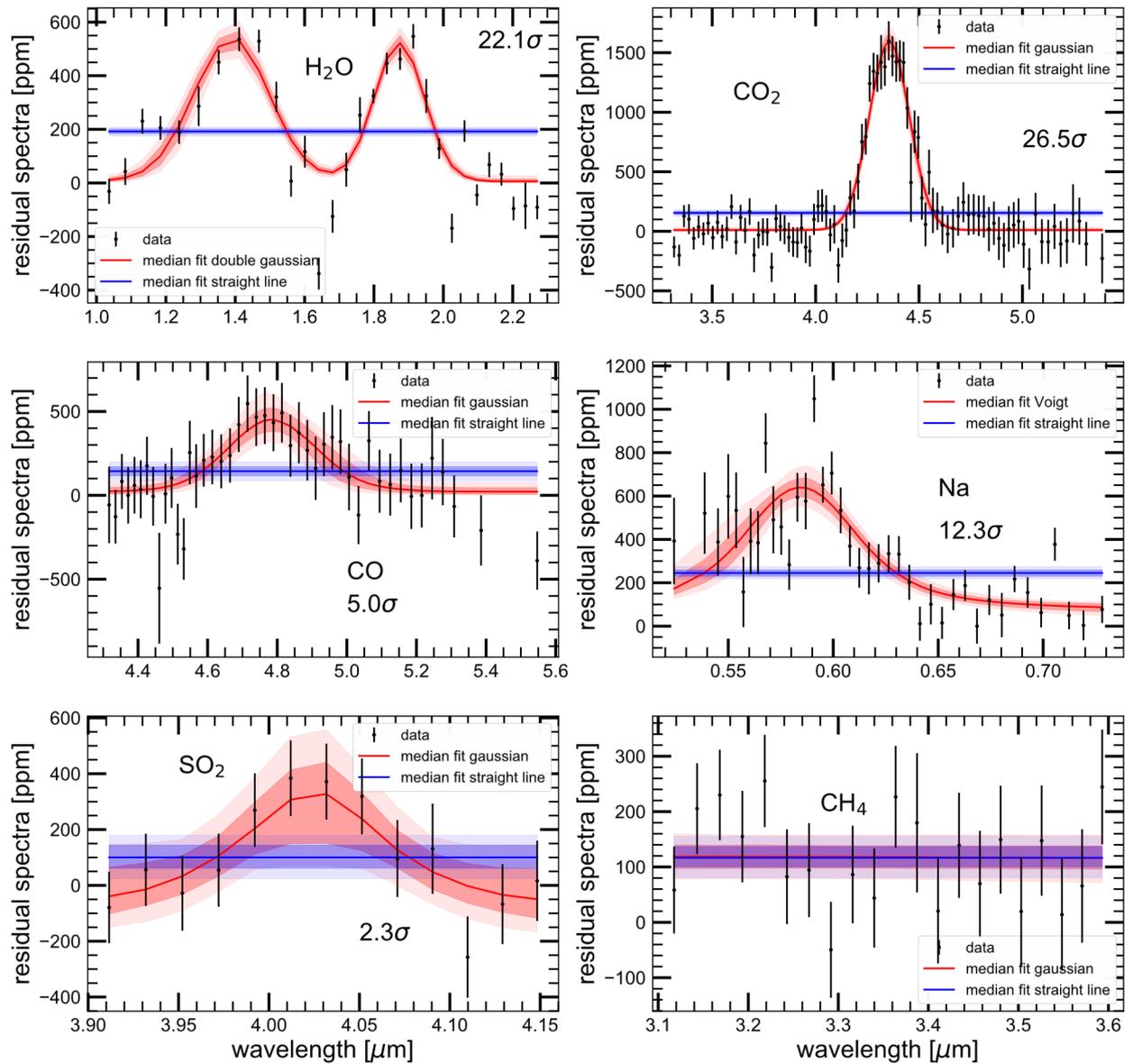

**Figure 12:** Each panel shows the residual spectrum of a particular gas. This residual spectrum was obtained by removing one gas at a time from the best-fit model atmosphere and subtracting the recalculated model transmission spectrum without that gas from the data. This residual spectrum was then fitted with a Gaussian distribution (and a Voigt profile for Na) and a constant offset, in a Bayesian framework. The median fit (solid lines) along with the $1\sigma$ and $2\sigma$ confidence intervals are shown with shaded red and blue regions for the Gaussian fits and the constant offset fits, respectively. The Bayes factor between the two functional fits was used to determine the detection significance of each gas. Note that the wavelength range covered in each panel is different.

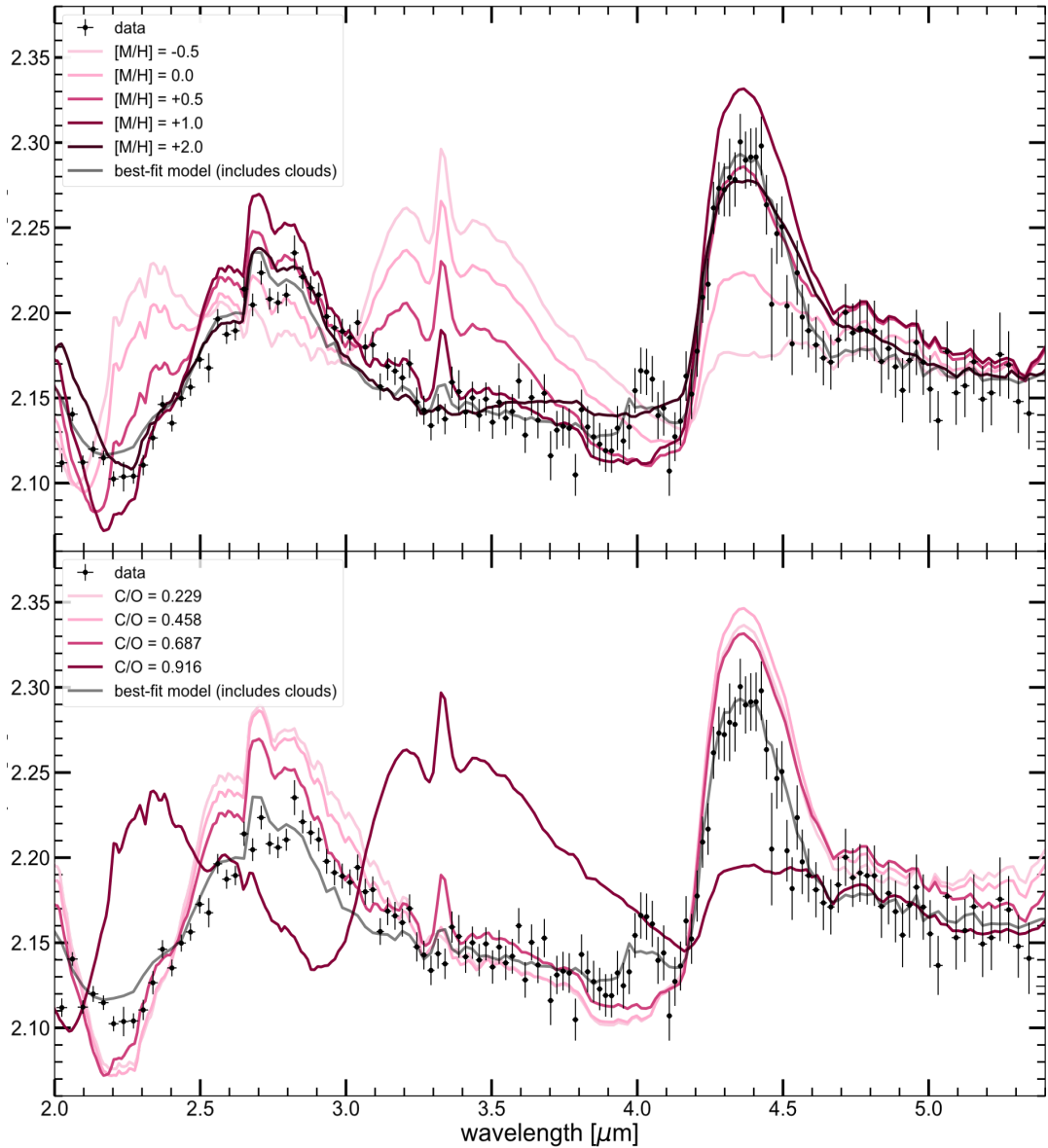

**Figure 13:** A comparison of cloud-free PICASO 3.0 RCTE models across a span of metallicities with the best-fit C/O ratio (0.68) is shown in the top panel. Each line coloured from faded to deep pink represents models with different metallicities between sub-solar to super-solar values. The simultaneous lack of a prominent $CH_4$ feature at 2.3 and 3.3 $\mu$m and the presence of a strong $CO_2$ feature indicate that the observations disfavor a low-metallicity atmosphere. The bottom panel shows transmission spectrum models with different C/O ratios from sub-solar to super-solar values at 10×solar metallicity compared with the observed spectrum. The cloudy best-fit model obtained with the grid retrieval framework also has been shown in both the panels with the grey line.

Table 4: Detection significances of individual opacity sources with our two techniques: Bayes factor analysis with gas removal, and Gaussian/Voigt fits to the residual absorption profiles. Note, a negative ln(B) indicates that that specific opacity source is not preferred by the data.

| Gas | Bayesian gas removal | | Residual fit | |
|---|---|---|---|---|
| | ln(B) | σ | ln(B) | σ |
| $H_2O$[75] | 537.7 | 32.9 | 242.3 | 22.1 |
| $CO_2$[76] | 374.3 | 27.5 | 348.9 | 26.5 |
| $CO$[77] | 24.6 | 7.3 | 10.68 | 5.0 |
| $H_2S$[78] | -44.9 | N/A | -0.4 | N/A |
| $CH_4$[79] | -8.9 | N/A | -0.05 | N/A |
| $SO_2$[80] | 2.2 | 2.7 | 1.6 | 2.3 |
| $Na$[81] | 173.9 | 18.8 | 73.2 | 12.3 |
| $K$[82] | 0.6 | 1.7 | -1.0 | N/A |
| Cloud | 209.8 | 20.6 | N/A | N/A |

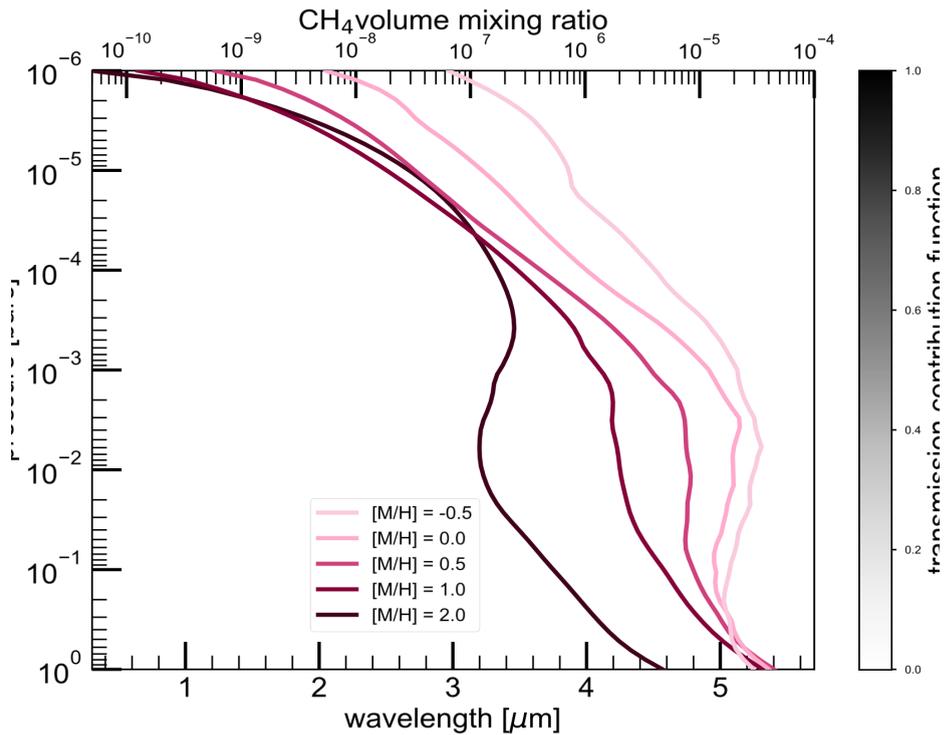

Fig. 14: The heat map shows the contribution function, as a function of wavelength in the lower x-axis and pressure in the y-axis, highlighting the parts of the atmosphere probed by the transmission observed data as a function of wavelength. The contribution function was calculated using the best-fit model. This shows that the data mostly probe pressure ranges between 0.1 to 2 mbars. The various shaded lines in pink show the volume mixing ratio of $CH_4$ (upper x-axis), from thermochemical equilibrium models, with different atmospheric metallicities at the best-fit C/O ratio of 0.68.

The best-fitting metallicity models can be used to place an upper limit on the $CH_4$ abundance, if the pressure ranges probed by the transmission spectrum are estimated. To estimate the pressure ranges probed by the data, we use the best-fit PICASO 3.0 model to calculate a pressure- and wavelength-dependent transmission contribution function of the atmosphere[86]. This contribution function for the best-fit 10×solar metallicity PICASO 3.0 model is shown as a heat-map in Fig. 14. This shows that the data mostly probes pressure ranges between 0.1–2 mbars. We also computed contribution functions for models with solar metallicity and find that they probe similar pressure ranges as well. Fig. 14 also shows the pressure dependent $CH_4$ abundances in models with different metallicities presented in Fig. 13 top panel. As only super-solar metallicity thermochemical equilibrium models are preferred by the data, the abundance profiles in Fig. 14 help us in putting an upper limit of $\lesssim 5\times10^{-6}$ on the $CH_4$ volume mixing ratio between 0.1–2 mbars.

**Clouds**

The observed spectrum shows somewhat muted transit depths, across the entire wavelength range, compared with the depths expected from clear atmospheric models. This hints toward some additional opacity source in the atmosphere with weak wavelength dependence. Opacity sources such as clouds can mute the spectral features in a transmission spectrum[2,4]. We post–process the transmission spectrum models with grey (i.e., wavelength-independent) cloud opacities to check whether they are preferred over clear atmospheric models by the data. However, the treatment of clouds differ between the four 1D RCTE model grids. PICASO 3.0 and ScCHIMERA grids implemented the cloud opacities using the following equation,

$$\tau_{i,cld} = \kappa_{cld}\frac{\delta P_i}{g} \quad (3)$$

where $\tau_{i,cld}$ is the cloud optical depth of the $i$'th atmospheric layer in the model with pressure–width $\delta P_i$ and $g$ represents the gravity of the planet. The best-fit value of the grey cloud opacity $\kappa_{cld} = 10^{-2.07}$ cm$^2$/g is calculated in a Bayesian framework by post-processing the RCTE model grid with this cloud opacity and comparing these post-processed models with the data. The ATMO grid includes grey cloud decks at multiple pressures between 1 and 50 mbars, but with variable factors 0, 0.5, 1, and 5 governing cloud opacity with respect to $H_2$'s scattering cross-section at 0.35 microns, where a factor 0 implies a cloud-free model spectrum. The PHOENIX grid includes similar cloud decks but between 0.3-10 mbars with cloud optical depth enhancement factors (identically defined as the ATMO grid) 0 and 10. We find that the cloudy models better fit the data than clear models across all four model grids. The contribution of clouds in limiting the depths of the gaseous features across the entire wavelength range is also shown in Fig. 4 with the grey shaded region.

**4 $\mu$m $SO_2$ feature identification**

None of the 1D RCTE models are able to capture the 4$\mu$m absorption feature seen in the data. We searched for multiple candidate gas species that could produce this feature if their abundances differ from the expected abundances from thermochemical equilibrium. The list of searched chemical species include C-bearing gases like $C_2H_2$, CS, $CS_2$, $C_2H_6$, $C_2H_4$, $CH_3$, CH, $C_2$, $CH_3Cl$, $CH_3F$, CN, and CP. Various metal hydrides, bromides, flourides and chlorides such as LiH, AlH, FeH, CrH, BeH, TiH, CaH, HBr, LiCl, HCl, HF, AlCl, NaF, and AlF were also searched as potential candidates to explain the feature. $SO_2$, $SO_3$, SO, and SH are among the sulphur-based gases which were considered. Other species which were considered include gases like $PH_3$, $H_2S$, HCN, $N_2O$, $GeH_4$, $SiH_4$, SiO, $AsH_3$, $H_2CO$, $H^+_3$, $OH^+$, KOH, Br$\alpha$-H, AlO, CN, CP, CaF, $H_2O_2$, $H_3O^+$, $HNO_3$, KF, MgO, PN, PO, PS, SiH, SiO2, SiS, TiO, and VO.

Among all these gases, $SO_2$ was the most promising candidate in terms of its spectral shape and chemical plausibility, although the expected chemical equilibrium abundance of $SO_2$ is too low to produce the absorption signal seen in the data. However, previous work exploring photochemistry in exoplanetary atmospheres[27,26] have shown that higher amounts of $SO_2$ can be created in the upper atmospheres of irradiated planets through photochemical processes. Therefore, we post-process the PICASO 3.0 and ScCHIMERA chemical equilibrium models with varying amounts of $SO_2$ in a Bayesian framework to estimate the $SO_2$ abundance required to explain the strength of the 4-$\mu$m feature. The required volume mixing ratio of $SO_2$ was found to be $\sim 10^{-5}$–$10^{-6}$. Note that in obtaining this estimate we assumed that the $SO_2$ volume mixing ratio does not vary with pressure for simplicity. In a photochemical scenario this assumption is likely not realistic, though the pressure ranged probed by $SO_2$ is also limited. Whether photochemical models can produce this amount of $SO_2$ in the atmospheric conditions of WASP-39 b is a pressing question which the ERS team is currently exploring (Welbanks et al. (in prep), Tsai et al. (in prep)). Whether this feature can be better explained by any other gaseous absorber is also currently under investigation by the ERS team.

**Data Availability**

The data used in this paper are associated with JWST program ERS 1366 and are available from the Mikulski Archive for Space Telescopes (https://mast.stsci.edu).

**Code Availability**

The codes used in this publication to extract, reduce, and analyse the data are as follows; STScI JWST

Calibration pipeline (https://github.com/spacetelescope/jwst), FIREFLy[24], tshirt[41], Eureka![48] (https://eurekadocs.readthedocs.io/en/latest/), and Tiberius[16,53,54].
In addition, these made use of Exoplanet[44](https://docs.exoplanet.codes/en/latest/), Pymc3[87] (https://docs.pymc.io/en/v3/index.html), ExoTEP [49,50,51], Batman[33] (http://lkreidberg.github.io/batman/docs/html/index.html), ExoTiC-ISM[55] (https://github.com/Exo-TiC/ExoTiC-ISM), ExoTiC-LD[56] (https://exotic-ld.readthedocs.io/en/latest/),
Emcee[32](https://emcee.readthedocs.io/en/stable/), Dynesty[70]

(https://dynesty.readthedocs.io/en/stable/index.html), and chromatic (https://zkbt.github.io/chromatic/),
which use the python libraries scipy[88], numpy[89], astropy[90,91], and matplotlib[92].
The atmospheric models used to fit the data can be found at PICASO[61,62,63,64] (https://natashabatalha.github.io/picaso/), Virga85 (https://natashabatalha.github.io/virga/), ScCHIMERA[59,60] (https://github.com/mrline/CHIMERA), ATMO[66,67], and PHOENIX[68,69].

## Acknowledgments

This work is based on observations made with the NASA/ESA/CSA JWST. The data were obtained from the Mikulski Archive for Space Telescopes at the Space Telescope Science Institute, which is operated by the Association of Universities for Research in Astronomy, Inc., under NASA contract NAS 5-03127 for JWST. These observations are associated with program JWST-ERS-01366. Support for program JWST-ERS-01366 was provided by NASA through a grant from the Space Telescope Science Institute, which is operated by the Association of Universities for Research in Astronomy, Inc., under NASA contract NAS 5-03127. This work benefited from the 2022 Exoplanet Summer Program in the Other Worlds Laboratory (OWL) at the University of California, Santa Cruz, a program funded by the Heising-Simons Foundation. We thank Eric Agol for constructive comments.

## Authors contributions

All authors played a significant role in one or more of the following: development of the original proposal, management of the project, definition of the target list and observation plan, analysis of the data, theoretical modeling, and preparation of this manuscript. Some specific contributions are listed as follows. provided overall program leadership and management. DS, EK, HW, IC, JLB, KBS, LK, MLM, MRL, NMB, VP, and ZBT made significant contributions to the design of the program. KBS generated the observing plan with input from the team. BB, EK, HW, IC, JLB, LK, MLM, MRL, NMB, and ZBT led or co-led working groups and/or contributed to significant strategic planning efforts like the design and implementation of the pre-launch Data Challenges. AC, DS, EvSc, NE, NG, TGP, VP generated simulated data for pre-launch testing of methods. ZR, DS, JK, EvSc, EM, AC reduced the data, modeled the light curves, and produced the planetary spectrum. SM, ML, JG, LJ generated theoretical model grids for comparison with the data. ZR, DS, CP, EM, EvSc, JK, ML, SM made significant contributions to the writing of this manuscript. ZR, DS, SM, EM, ML generated figures for this manuscript. BR, DP, IC, JT, JG, JB, MLM, RH, RM, SEM, TD contributed to the writing of this manuscript.

**Methods References**

**Acknowledgments**

This work is based on observations made with the NASA/ESA/CSA JWST. The data were obtained from the Mikulski Archive for Space Telescopes at the Space Telescope Science Institute, which is operated by the Association of Universities for Research in Astronomy, Inc., under NASA contract NAS 5-03127 for JWST. These observations are associated with program JWST-ERS-01366. Support for program JWST-ERS-01366 was provided by NASA through a grant from the Space Telescope Science Institute, which is operated by the Association of Universities for Research in Astronomy, Inc., under NASA contract NAS 5-03127. This work benefited from the 2022 Exoplanet Summer Program in the Other Worlds Laboratory (OWL) at the University of California, Santa Cruz, a program funded by the Heising-Simons Foundation. We thank Eric Agol for constructive comments.

**Author contributions**

All authors played a significant role in one or more of the following: development of the original proposal, management of the project, definition of the target list and observation plan, analysis of the data, theoretical modeling, and preparation of this manuscript. Some specific contributions are listed as follows. provided overall program leadership and management. DS, EK, HW, IC, JLB, KBS, LK, MLM, MRL, NMB, VP, and ZBT made significant contributions to the design of the program. KBS generated the observing plan with input from the team. BB, EK, HW, IC, JLB, LK, MLM, MRL, NMB, and ZBT led or co-led working groups and/or contributed to significant strategic planning efforts like the design and implementation of the pre-launch Data Challenges. AC, DS, EvSc, NE, NG, TGP, VP generated simulated data for pre-launch testing of methods. ZR, DS, JK, EvSc, EM, AC reduced the data, modeled the light curves, and produced the planetary spectrum. SM, ML, JG, LJ generated theoretical model grids for comparison with the data. ZR, DS, CP, EM, EvSc, JK, ML, SM made significant contributions to the writing of this manuscript. ZR, DS, SM, EM, ML generated figures for this manuscript. BR, DP, IC, JT, JG, JB, MLM, RH, RM, SEM, TD contributed to the writing of this manuscript.

**Author Affiliations**

[1] Department of Earth and Planetary Sciences, Johns Hopkins University, Baltimore, Maryland, USA
[2] Department of Physics & Astronomy, Johns Hopkins University, Baltimore, MD, USA
[3] Department of Astronomy & Astrophysics, University of California, Santa Cruz, Santa Cruz, CA, 95064, USA
[4] Johns Hopkins APL, Laurel, MD, USA
[5] Center for Astrophysics | Harvard & Smithsonian, Cambridge, MA 02138, USA
[6] Department of Physics, Imperial College London, London, UK

[7] Imperial College Research Fellow
[8] Steward Observatory, University of Arizona, Tucson, AZ, USA
[9] School of Earth & Space Exploration, Arizona State University, Tempe, AZ, USA
[10] Institut de Recherche sur les Exoplanètes, Département de Physique, Université de Montréal, Montreal, Canada
[11] NASA Ames Research Center, Moffett Field, CA, USA
[12] School of Earth and Planetary Sciences (SEPS), National Institute of Science Education and Research (NISER), HBNI, Jatani, Odisha, India
[13] Department of Physics, Utah Valley University, Orem, UT, USA
[14] Department of Astronomy, University of Michigan, Ann Arbor, MI, USA
[15] Department of Astronomy and Carl Sagan Institute, Cornell University, Ithaca, NY, USA
[16] NHFP Sagan Fellow
[17] Lunar and Planetary Laboratory, University of Arizona, Tucson, AZ, USA
[18] School of Physics, University of Bristol, HH Wills Physics Laboratory, Tyndall Avenue, Bristol, BS8 1TL, UK
[19] Space Telescope Science Institute, 3700 San Martin Dr, Baltimore, MD 21218, USA
[20] Department of Astronomy & Astrophysics, University of Chicago, Chicago, IL, USA
[21] Department of Physics and Institute for Research on Exoplanets, Université de Montréal, Montreal, QC, Canada
[22] Department of Astrophysical and Planetary Sciences, University of Colorado, Boulder, CO, USA
[23] Department of Physics and Astronomy, University of Kansas, Lawrence, KS, USA
[24] Earth & Planets Laboratory, Carnegie Institution of Washington, Washington, DC, USA
[25] Max Planck Institute for Astronomy, Heidelberg, Germany
[26] Harvard & Smithsonian, Center for Astrophysics, Cambridge, MA, USA.
[27] NASA Sagan Fellow
[28] INAF - Osservatorio Astrofisico di Torino
[29] School of Physics, Trinity College Dublin, Dublin, Ireland
[30] Laboratoire d'Astrophysique de Bordeaux, CNRS, Université de Bordeaux, Pessac, France
[31] University Observatory Munich, Ludwig Maximilian University, Munich, Bavaria, Germany
[32] Exzellenzcluster Origins, Garching, Germany
[33] Université Côte d'Azur, Observatoire de la Côte d'Azur, CNRS, Laboratoire Lagrange, Frabce
[34] Atmospheric, Oceanic and Planetary Physics, Department of Physics, University of Oxford, Oxford, UK
[35] Department of Physics and Institute for Research on Exoplanets, Université de Montréal, Montreal, QC, Canada
[36] Department of Physics, University of Oxford, Oxford, UK
[37] Department of Astronomy and Carl Sagan Institute, Cornell University, Ithaca, NY, USA
[38] NHFP Sagan Fellow
[39] Centre for Exoplanets and Habitability, University of Warwick, Coventry, UK
[40] Department of Physics, University of Warwick, Coventry, UK
[41] Indian Institute of Technology Indore, India
[42] NSF Graduate Research Fellow

[43] School of Physical Sciences, The Open University, Milton Keynes, UK
[44] Anton Pannekoek Institute for Astronomy, University of Amsterdam, Amsterdam, The Netherlands
[45] Centro de Astrobiología (CSIC-INTA), ESAC Campus, Camino Bajo del Castillo s/n, 28692 Villanueva de la Cañada, Madrid
[46] BAER Institute, NASA Ames Research Center, Moffet Field, Mountain View, CA, USA
[47] New York University Abu Dhabi, Abu Dhabi, United Arab Emirates
[48] Center for Astro, Particle and Planetary Physics (CAP3), New York University Abu Dhabi, Abu Dhabi, UAE
[49] Department of Physics & Astronomy, University of Kansas, Lawrence, KS, USA
[50] School of Physics and Astronomy, University of Leicester, University Road, Leicester, LE1 7RH
[51] European Space Agency (ESA), ESA Baltimore Office, Baltimore MD, United States of America.
[52] Department of Physics and Astronomy, University College London, London, United Kingdom.
[53] Centre for Exoplanet Science, University of St Andrews, St Andrews, UK
[54] Leiden Observatory, Leiden University, Leiden, Zuid-Holland, the Netherlands
[55] Department of Astrophysical Sciences, Princeton University, 4 Ivy Lane, Princeton, NJ 08544
[56] LSSTC Catalyst Fellow
[57] Department of Physics and Astronomy, KU Leuven, Leuven, Belgium
[58] Department of Astronomy and Carl Sagan Institute, Cornell University, Ithaca, NY, USA
[59] Planetary Science Group, Department of Physics and Florida Space Institute, University of Central Florida, Orlando, Florida, USA
[60] Jet Propulsion Laboratory, California Institute of Technology, CA, USA
[61] Division of Geological and Planetary Sciences, California Institute of Technology, CA, USA
[62] Institute for Astrophysics, University of Vienna, Vienna, Austria
[63] Jet Propulsion Laboratory, California Institute of Technology, Pasadena, CA 91125 USA
[64] Department of Astronomy, University of Maryland, College Park, MD, USA
[65] California Institute of Technology, IPAC, Pasadena, CA, 91125, USA
[66] Département d'astronomie, Université de Genève, Geneva, Switzerland
[67] Centro de Astrobiología (CAB), INTA-CSIC, ESAC campus, 28692, Villanueva de la Cañada (Madrid), Spain
[68] Department of Physics, University of Rome ``Tor Vergata'', Rome, Italy
[69] INAF - Osservatorio Astrofisico di Torino, Italy
[70] Department of Physics and Astronomy, Faculty of Environment, Science and Economy
[71] SRON Netherlands Institute for Space Research, Leiden, The Netherlands
[72] Instituto de Astrofisica de Canarias (IAC), La Laguna, Tenerife, Spain
[73] Departamento de Astrofisica, Universidad de La Laguna (ULL), La Laguna, Tenerife, Spain;
[74] INAF ‚Äì Palermo Astronomical Observatory, Palermo, Italy
[75] Department of Earth, Atmospheric and Planetary Sciences, Massachusetts Institute of Technology, Cambridge, MA, USA

[76] Kavli Institute for Astrophysics and Space Research, Massachusetts Institute of Technology, Cambridge, MA, USA
[77] 51 Pegai b Fellow
[78] Astronomy Department and Van Vleck Observatory, Wesleyan University, Middletown, CT 06459, USA
[79] Institute of Astronomy, University of Cambridge, United Kingdom
[80] Astrophysics Group, Keele University, Staffordshire, UK
[81] Université Paris-Saclay, UVSQ, CNRS, CEA, Maison de la Simulation, 91191, Gif-sur-Yvette, France.
[82] Department of Physics, Brown University, Providence, RI, USA
[83] Université de Paris Cité and Univ Paris Est Creteil, CNRS, LISA, F-75013 Paris, France
[84] Astrophysics and Planetary Sciences, University of Colorado Boulder, Boulder, CO, USA
[85] Department of Earth and Planetary Sciences, University of California Santa Cruz, Santa Cruz, California, USA